%% file: svFit4tau.tex
\def\verPreprint{1}
\def\verPAPER{2}
\def\ver{1}

\ifx\ver\verPreprint
\documentclass[a4paper,english,11pt]{article}
\usepackage[bindingoffset=0.5cm,left=2.5cm,right=2.5cm,top=2.5cm,bottom=2.5cm,footskip=1.0cm]{geometry}
\usepackage{lineno,hepnames,bm,multirow,amssymb,authblk,graphicx,newclude,xspace,hyperref}
\fi
\ifx\ver\verPAPER
\documentclass[1p]{elsarticle}
\usepackage{lineno,hyperref,hepnames,bm,multirow,amssymb,xspace}
\fi

\renewcommand{\Plepton}{\ensuremath{\ell}}
\newcommand{\Phadron}{\ensuremath{\textrm{h}}}
\newcommand{\tauh}{\ensuremath{\Pgt_{\textrm{h}}}\xspace}
\newcommand{\tauhnu}{\ensuremath{\tauh \, \Pnu_{\kern-0.10em \Pgt}}\xspace}
\newcommand{\tauhnuOne}{\ensuremath{\tauh^{(1)} \, \Pnu^{(1)}_{\kern-0.10em \Pgt}}\xspace}
\newcommand{\tauhnuTwo}{\ensuremath{\tauh^{(2)} \, \Pnu^{(2)}_{\kern-0.10em \Pgt}}\xspace}
\newcommand{\ellnunu}{\ensuremath{\Plepton \, \APnu_{\kern-0.10em \Plepton} \, \Pnu_{\kern-0.10em \Pgt}}\xspace}
\newcommand{\ellnunuOne}{\ensuremath{\Plepton^{(1)} \, \APnu^{(1)}_{\kern-0.10em \Plepton} \, \Pnu^{(1)}_{\kern-0.10em \Pgt}}\xspace}
\newcommand{\ellnunuTwo}{\ensuremath{\Plepton^{(2)} \, \APnu^{(2)}_{\kern-0.10em \Plepton} \, \Pnu^{(2)}_{\kern-0.10em \Pgt}}\xspace}
\newcommand{\ellMinusnunu}{\ensuremath{\Plepton^{-} \, \APnu_{\kern-0.10em \Plepton} \, \Pnu_{\kern-0.10em \Pgt}}\xspace}
\newcommand{\ellPlusnunu}{\ensuremath{\Plepton^{+} \, \Pnu_{\kern-0.10em \Plepton} \, \APnu_{\kern-0.10em \Pgt}}\xspace}
\newcommand{\enunu}{\ensuremath{\Pe \, \APnu_{\kern-0.10em \Pe} \, \Pnu_{\kern-0.10em \Pgt}}\xspace}
\newcommand{\mununu}{\ensuremath{\Pgm \, \APnu_{\kern-0.10em \Pgm} \, \Pnu_{\kern-0.10em \Pgt}}\xspace}

\newcommand{\pT}{\ensuremath{p_{\textrm{T}}}\xspace}
\newcommand{\pThat}{\ensuremath{\hat{p}_{\textrm{T}}}\xspace}

\newcommand{\pX}{\ensuremath{p_{\textrm{x}}}\xspace}
\newcommand{\pXhat}{\ensuremath{\hat{p}_{\textrm{x}}}\xspace}
\newcommand{\pY}{\ensuremath{p_{\textrm{y}}}\xspace}
\newcommand{\pYhat}{\ensuremath{\hat{p}_{\textrm{y}}}\xspace}

\newcommand{\MET}{\ensuremath{p_{\textrm{T}}^{\textrm{\kern0.10em miss}}}\xspace}
\newcommand{\vecMET}{\ensuremath{\bm{p}_{\textrm{T}}^{\textrm{\kern0.10em miss}}}\xspace}
\newcommand{\METx}{\ensuremath{p_{\textrm{x}}^{\textrm{\kern0.10em miss}}}\xspace}
\newcommand{\METy}{\ensuremath{p_{\textrm{y}}^{\textrm{\kern0.10em miss}}}\xspace}

\newcommand{\GeV}{\ensuremath{\textrm{GeV}}\xspace}
\newcommand{\TeV}{\ensuremath{\textrm{TeV}}\xspace}
\newcommand{\rec}{\ensuremath{\textrm{rec}}}

\newcommand{\vis}{\ensuremath{\textrm{vis}}}
\newcommand{\inv}{\ensuremath{\textrm{inv}}}
\newcommand{\eff}{\ensuremath{\textrm{eff}}}

\newcommand{\BW}{\ensuremath{\textrm{BW}}}

\usepackage{array}
\newcolumntype{C}[1]{>{\centering\arraybackslash}p{#1}}

\begin{document}

\ifx\ver\verPAPER
\begin{frontmatter}
\fi

\title{Reconstruction of the mass of Higgs boson pairs in events with Higgs boson pairs
  decaying into four $\Pgt$ leptons}


\ifx\ver\verPreprint
\author[1]{Karl Ehat\"aht}
\author[1]{Luca Marzola}
\author[1]{Christian Veelken}
\affil[1]{National Institute for Chemical Physics and Biophysics, 10143 Tallinn, Estonia}
\fi
\ifx\ver\verPAPER
\author[tallinn]{Karl Ehat\"aht}
\ead{karl.ehataht@cern.ch}
\author[tartu]{Luca Marzola}
\ead{luca.marzola@ut.ee}
\author[tallinn]{Christian Veelken}
\ead{christian.veelken@cern.ch}
\address[tallinn]{National Institute for Chemical Physics and Biophysics, 10143 Tallinn, Estonia}
\fi

\ifx\ver\verPreprint
\maketitle
\fi

\begin{abstract}
Various theories beyond the Standard Model predict the existence of heavy resonances decaying to Higgs ($\PHiggs$) boson pairs.
In order to maximize the sensitivity of searches for such resonances, 
it is important that experimental analyses cover a variety of decay modes.
The decay of $\PHiggs$ boson pairs to four $\Pgt$ leptons ($\PHiggs\PHiggs \to \Pgt\Pgt\Pgt\Pgt$) has not been discussed in the literature so far.
This decay mode provides a small branching fraction, but also comparatively low backgrounds.
We present an algorithm for the reconstruction of the mass of the $\PHiggs$ boson pair in events in which the $\PHiggs$ boson pair
decays via $\PHiggs\PHiggs \to \Pgt\Pgt\Pgt\Pgt$ and the $\Pgt$ leptons subsequently decay into electrons, muons, or hadrons.
The algorithm achieves a resolution of $7$--$22\%$ relative to the mass of the $\PHiggs$ boson pair, 
depending on the mass of the resonance.
\end{abstract}

\ifx\ver\verPAPER
\end{frontmatter}
\fi

\clearpage


\include*{introduction}
\include*{algorithm}

\include*{performance}
\include*{summary}

\include*{appendix}

\bibliography{svFit4tau}

\end{document}

%% file: introduction.tex
\section{Introduction}
\label{sec:introduction}

The discovery of the Higgs ($\PHiggs$) boson by the ATLAS and CMS experiments at the LHC~\cite{Higgs-Discovery_CMS,Higgs-Discovery_ATLAS}
represents a major step towards our understanding of electroweak symmetry breaking (EWSB) 
and of the mechanism that generates the masses of quarks and leptons, which constitute the ``ordinary'' matter in our Universe.
In a combined analysis of the data recorded by ATLAS and CMS, 
the mass, $m_{\PHiggs}$, of the $\PHiggs$ boson has been measured to be $m_{\PHiggs} = 125.09 \pm 0.24$~\GeV~\cite{HIG-14-042}.
The Standard Model (SM) of particle physics makes precise predictions for all properties of the $\PHiggs$ boson, 
given its mass and the vacuum expectation value $v=246$~\GeV~\cite{PDG} of the Higgs field.
So far, all properties that have been measured agree with the expectation for a SM $\PHiggs$ boson~\cite{HIG-15-002}.
The rate for its decay to a pair of $\Pgt$ leptons has been measured recently
and found to be consistent with the SM expectation within the uncertainties of these measurements, 
at present amounting to $20$--$30\%$~\cite{HIG-13-004,Aad:2015vsa,HIG-15-002,HIG-16-043,ATLAS:2018lur},
One important prediction of the SM yet has to be verified experimentally, however:
the $\PHiggs$ boson self-interaction.

\begin{figure}[h!]
\setlength{\unitlength}{1mm}
\begin{center}
\begin{picture}(180,34)(0,0)
\put(1.5, 0.0){\mbox{\includegraphics*[height=34mm]
  {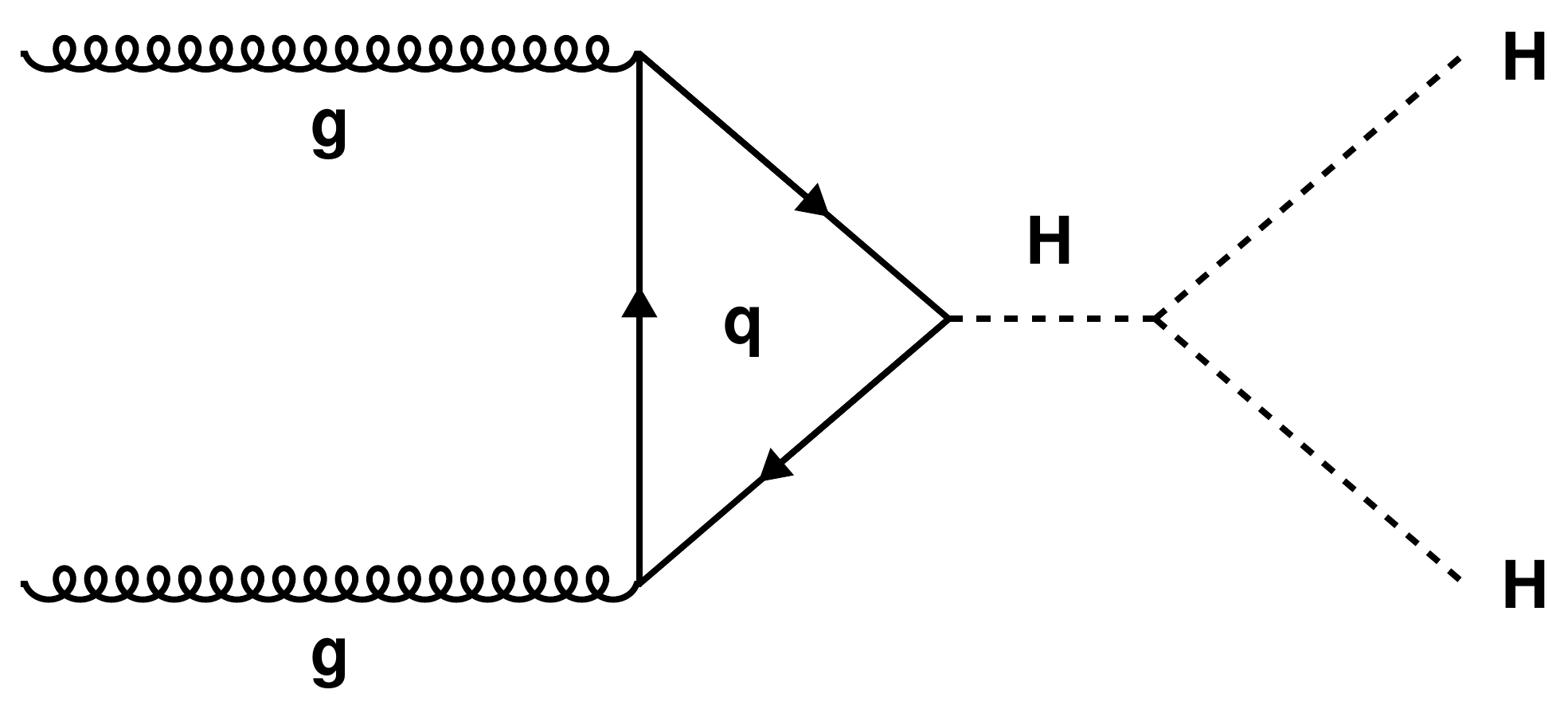}}}
\put(81.5, 0.0){\mbox{\includegraphics*[height=34mm]
  {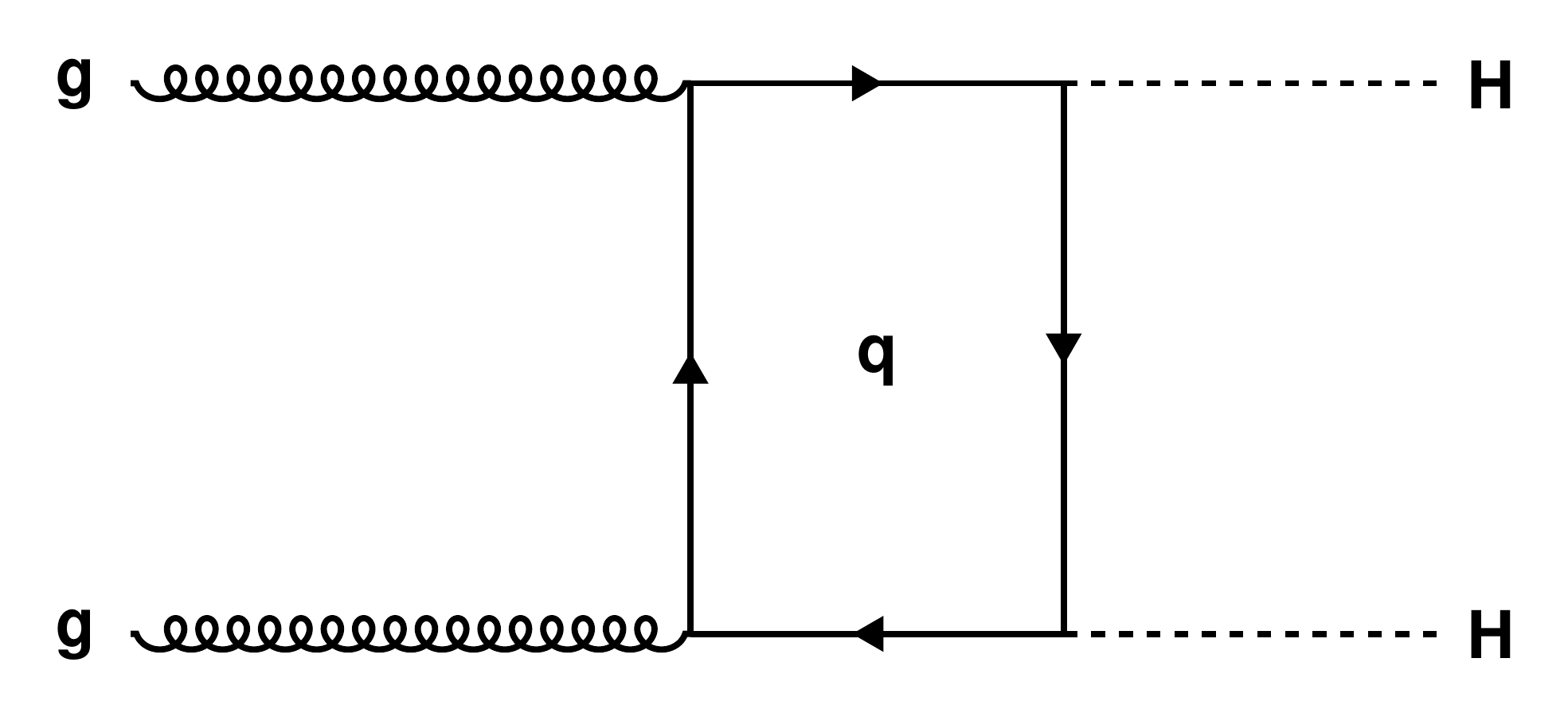}}}
\end{picture}
\end{center}
\caption{ LO Feynman diagrams for $\PHiggs\PHiggs$ production within the SM.}
\label{fig:FeynmanDiagrams_smHH}
\label{fig:massDistributions}
\end{figure}

The SM predicts $\PHiggs$ boson self-interactions via trilinear and quartic couplings. 
Measurements of the $\PHiggs$ boson self-interactions will ultimately either confirm or falsify 
that the Brout-Englert-Higgs mechanism of the SM is responsible for EWSB
and the flavour hierarchy of the SM. 
The trilinear coupling ($\lambda_{\PHiggs\PHiggs\PHiggs}$) can be determined at the LHC, 
by measuring the rate for $\PHiggs$ boson pair ($\PHiggs\PHiggs$) production. 
The measurement is challenging, because of the small signal cross section, 
which results from the destructive interference of two competing production processes, and suffers from sizeable backgrounds. 
The leading order (LO) Feynman diagrams for SM $\PHiggs\PHiggs$ production are shown in Fig.~\ref{fig:FeynmanDiagrams_smHH}.
The cross section amounts to about $\sigma = 34$~fb in proton-proton collisions at $\sqrt{s}=13$~\TeV center-of-mass energy.
The ``triangle'' diagram shown on the left depends on $\lambda_{\PHiggs\PHiggs\PHiggs}$,
while the ``box'' diagram shown on the right does not.
The quartic coupling is not accessible at the LHC, as the cross section of the corresponding process, 
triple $\PHiggs$ boson production, is too small to be measured
even with the large dataset that is expected to be collected by the end of the LHC operation.

Deviations of $\lambda_{\PHiggs\PHiggs\PHiggs}$ from its SM value of $\lambda_{\textrm{SM}}=\frac{m_{\PHiggs}^2}{2 \, v}$, referred to as anomalous $\PHiggs$ boson self-couplings,
alter the interference between the two diagrams, 
resulting in a change in the $\PHiggs\PHiggs$ production cross section and a change in the distribution of the mass, $m_{\PHiggs\PHiggs}$, of the $\PHiggs$ boson pair.
Regardless of the value of $\lambda_{\PHiggs\PHiggs\PHiggs}$, a broad distribution in $m_{\PHiggs\PHiggs}$ is expected,
motivating the convention to refer to the interference of box and triangle diagram as ``non-resonant'' $\PHiggs\PHiggs$ production.
The shape of the distribution in $m_{\PHiggs\PHiggs}$ provides a handle to determine $\lambda_{\PHiggs\PHiggs\PHiggs}$,
complementary to measuring the $\PHiggs\PHiggs$ production cross section.
Various scenarios beyond the SM feature anomalous $\PHiggs$ boson self-couplings,
for example two-Higgs-doublet models~\cite{Branco:2011iw}, the minimal supersymmetric extension of the SM (MSSM)~\cite{Gunion:1989we},
and models with composite $\PHiggs$ bosons~\cite{Grober:2010yv,Contino:2012xk}.
The prospects for improving the sensitivity to determine $\lambda_{\PHiggs\PHiggs\PHiggs}$ by utilising information on the mass of the $\PHiggs$ boson pair
have been studied in events in which the $\PHiggs$ boson pair decays via $\PHiggs\PHiggs \to \PW\PW\PW\PW$, with subsequent decay of the $\PW$ bosons to electrons, muons, or jets,
in Refs.~\cite{Baur:2002rb,Baur:2002qd}.
The information that can be extracted from the distribution in $m_{\PHiggs\PHiggs}$ is limited, however,
by the fact that the distribution in $m_{\PHiggs\PHiggs}$ changes only moderately with $\lambda_{\PHiggs\PHiggs\PHiggs}$.

The rate for $\PHiggs\PHiggs$ production may be enhanced significantly in case an as yet undiscovered heavy resonance $\textrm{X}$ decays into pairs of $\PHiggs$ bosons.
Several models beyond the SM give rise to such decays, 
for example Higgs portal models~\cite{Englert:2011yb,No:2013wsa} and models involving warped extra dimensions~\cite{Randall:1999ee},
as well as two-Higgs-doublet models and models with composite Higgs bosons.
If the lifetime $t$ of the resonance is sufficiently large, $t \gtrsim 10^{-25}\textrm{~s}/m_{\textrm{X}}\textrm{~[100~GeV]}$, 
where $m_{\textrm{X}}$ denotes the mass of the resonance, 
the distribution in $m_{\PHiggs\PHiggs}$ is expected to exhibit a narrow peak at $m_{\textrm{X}}$.

In this paper we present an algorithm for reconstructing the mass $m_{\PHiggs\PHiggs}$ of the $\PHiggs$ boson pair 
in events in which the $\PHiggs$ boson pair originates from the decay of a heavy resonance $\textrm{X}$ 
and decays via $\PHiggs\PHiggs \to \Pgt\Pgt\Pgt\Pgt$,
with subsequent decay of the $\Pgt$ leptons via $\Pgt \to \enunu$, $\Pgt \to \mununu$, or $\Pgt \to \textrm{hadrons} + \Pnut$.
We refer to $\Pgt$ decays to an electron or muon (to hadrons) as ``leptonic'' (``hadronic'') $\Pgt$ decays.
The system of hadrons produced in a hadronic $\Pgt$ decay is denoted by the symbol $\tauh$.
The decay of $\PHiggs$ boson pairs to four $\Pgt$ leptons ($\PHiggs\PHiggs \to \Pgt\Pgt\Pgt\Pgt$) has not been discussed in the literature so far.
This decay channel provides a small branching fraction, but is expected to benefit from comparatively low backgrounds.
The resolution on $m_{\PHiggs\PHiggs}$, achieved by our algorithm, varies between $7$ and $22\%$,
depending on the mass of the resonance.
We expect that the reconstruction of $m_{\PHiggs\PHiggs}$ will significantly improve 
the separation of the $\PHiggs\PHiggs \to \Pgt\Pgt\Pgt\Pgt$ signal from residual backgrounds,
thereby increasing the sensitivity to either find evidence for the presence of such a signal in the LHC data or to set stringent exclusion limits.

The reconstruction of $m_{\PHiggs\PHiggs}$ in $\PHiggs\PHiggs \to \Pgt\Pgt\Pgt\Pgt$ events is based on the formalism,
developed in Ref.~\cite{SVfitMEM}, for treating $\Pgt$ lepton decays in the so-called matrix element (ME) method~\cite{Kondo:1988yd,Kondo:1991dw}.
The algorithm presented in this paper does not employ the full ME treatment,
but is based on a simplified likelihood approach.
The simplified approach is motivated by the studies performed in Ref.~\cite{SVfitMEM}, 
which found that the difference in mass resolution between the approximate likelihood treatment and the full ME formalism,
applied to the task of reconstructing the mass of the $\PHiggs$ boson in events containing a single $\PHiggs$ boson that decays via $\PHiggs \to \Pgt\Pgt$,
is small, while the likelihood approach provides a significant reduction in computing time.

Our algorithm for reconstructing the mass of the $\PHiggs$ boson pair 
in $\PHiggs\PHiggs \to \Pgt\Pgt\Pgt\Pgt$ events is presented in Section~\ref{sec:algorithm}.
The resolution achieved by the algorithm in reconstructing $m_{\PHiggs\PHiggs}$ 
for hypothetical resonances $\textrm{X}$ of different mass is studied in Section~\ref{sec:performance}.
The paper concludes with a summary in Section~\ref{sec:summary}.

%% file: algorithm.tex
\section{The algorithm}
\label{sec:algorithm}

The reconstruction of the mass of the $\PHiggs$ boson pair is based on maximizing the likelihood function:
\begin{align}
& \mathcal{P}(\bm{p}^{\vis(1)},\bm{p}^{\vis(2)},\bm{p}^{\vis(3)},\bm{p}^{\vis(4)};\pX^{\rec},\pY^{\rec}|m_{\textrm{X}})
= \frac{32\pi^{4}}{s} \, \int \, d\Phi_{n} \, \cdot \nonumber \\
& \quad \delta\left( \left(\sum_{i=1}^{4} \, \hat{E}_{\Pgt(i)}\right)^{2} - \left(\sum_{i=1}^{4} \, \bm{\hat{p}}^{\Pgt(i)}\right)^{2} - m_{\textrm{X}} \right) 
  \delta\left( \pXhat^{\rec} + \sum_{i=1}^{4} \pXhat^{\Pgt(i)} \right) \cdot \delta\left( \pYhat^{\rec} + \sum_{i=1}^{4} \pYhat^{\Pgt(i)} \right) \cdot \nonumber \\
& \quad \vert \BW^{(1)}_{\Pgt} \vert^{2} \cdot \vert \mathcal{M}^{(1)}_{\Pgt\to\cdots}(\bm{\hat{p}}) \vert^{2} \cdot W(\bm{p}^{\vis(1)}|\bm{\hat{p}}^{\vis(1)}) \, \cdot 
\, \vert \BW^{(2)}_{\Pgt} \vert^{2} \cdot \vert \mathcal{M}^{(2)}_{\Pgt\to\cdots}(\bm{\hat{p}}) \vert^{2} \cdot W(\bm{p}^{\vis(2)}|\bm{\hat{p}}^{\vis(2)}) \, \cdot \nonumber \\
& \quad \vert \BW^{(3)}_{\Pgt} \vert^{2} \cdot \vert \mathcal{M}^{(3)}_{\Pgt\to\cdots}(\bm{\hat{p}}) \vert^{2} \cdot W(\bm{p}^{\vis(3)}|\bm{\hat{p}}^{\vis(3)}) \, \cdot
\, \vert \BW^{(4)}_{\Pgt} \vert^{2} \cdot \vert \mathcal{M}^{(4)}_{\Pgt\to\cdots}(\bm{\hat{p}}) \vert^{2} \cdot W(\bm{p}^{\vis(4)}|\bm{\hat{p}}^{\vis(4)}) \, \cdot \nonumber \\
& \quad W_{\rec}( \pX^{\rec},\pY^{\rec} | \pXhat^{\rec},\pYhat^{\rec} ) 
\label{eq:likelihood}
\end{align}
with respect to the parameter $m_{\textrm{X}}$, 
the mass of the postulated heavy particle $\textrm{X}$ that decays into a pair of $\PHiggs$ bosons.
We refer to the electron, muon, or hadrons produced in each $\Pgt$ decay as the ``visible'' $\Pgt$ decay products.
Their energy (momentum) is denoted by the symbol $E_{\vis(i)}$ ($\bm{p}^{\vis(i)}$), where the index $i$ ranges between $1$ and $4$.
The symbol $E_{\Pgt(i)}$ ($\bm{p}^{\Pgt(i)}$) denotes the energy (momentum) of the $i$-th $\Pgt$ lepton.
Bold letters represent vector quantities.
The true values of energies and momenta are indicated by a hat,
while symbols without a hat represent the measured values.
We use a Cartesian coordinate system, the $z$-axis of which is defined by the proton beam direction.
The symbol $d\Phi_{n} = \prod_{i}^{n} \,
\frac{d^{3}\bm{p}^{(i)}}{(2\pi)^{3} \, 2 E_{(i)}}$ denotes the differential $n$-particle phase-space element,
where $n$ refers to the number of particles in the final state.
The symbol $\vert \BW^{(i)}_{\Pgt} \vert^{2} \cdot \vert \mathcal{M}^{(i)}_{\Pgt\to\cdots}(\bm{\hat{p}}) \vert^{2}$ 
denotes the squared modulus of the ME for the decay of the $i$-th $\Pgt$ lepton.
The $\delta$-function 
$\delta\left( \left(\sum_{i=1}^{4} \, E_{\Pgt(i)}\right)^{2} - \left(\sum_{i=1}^{4} \, \bm{\hat{p}}^{\Pgt(i)}\right)^{2} - m_{\textrm{X}} \right)$
enforces the condition that the mass of the system of four $\Pgt$ leptons equals the value of the parameter $m_{\textrm{X}}$
given on the left-hand-side of the equation.

The functions $W(\bm{p}^{\vis(i)}|\bm{\hat{p}}^{\vis(i)})$ and $W_{\rec}( \pX^{\rec},\pY^{\rec} | \pXhat^{\rec},\pYhat^{\rec} )$ are referred to as ``transfer functions'' (TF).
They quantify the experimental resolutions with which the momenta of particles are measured in the detector.
The nomenclature $W(\bm{p}|\bm{\hat{p}})$ has the following meaning:
The value of the function $W(\bm{p}|\bm{\hat{p}})$ represents the probability density to observe the measured momentum $\bm{p}$,
given that the true value of the momentum is $\bm{\hat{p}}$.
The function $W(\bm{p}^{\vis(i)}|\bm{\hat{p}}^{\vis(i)})$ represents the resolution for measuring the momentum of the visible $\Pgt$ decay products,
while the function $W_{\rec}( \pX^{\rec},\pY^{\rec} | \pXhat^{\rec},\pYhat^{\rec} )$ quantifies the resolution for measuring the momentum, 
in the $x$-$y$ plane, of the hadronic recoil.

The hadronic recoil is defined as the vectorial sum of all particles in the event that do not originate from the decay of the two $\PHiggs$ bosons.
Conservation of momentum in the plane transverse to the beam direction implies that
the components $\pXhat^{\rec}$ and $\pYhat^{\rec}$ of its true momentum are equal to the negative sum of the momentum components
$\pXhat^{\Pgt(i)}$ and $\pYhat^{\Pgt(i)}$ of the four $\Pgt$ leptons,
\begin{equation*}
\pXhat^{\rec} = -\left( \sum_{i=1}^{4} \, \pXhat^{\Pgt(i)} \right) \quad \mbox{ and } \quad \pYhat^{\rec} = -\left( \sum_{i=1}^{4} \, \pYhat^{\Pgt(i)} \right) \, ,
\end{equation*}
as enforced by the two $\delta$-functions
$\delta\left( \pXhat^{\rec} + \sum_{i=1}^{4} \pXhat^{\Pgt(i)} \right)$ and $\delta\left( \pYhat^{\rec} + \sum_{i=1}^{4} \pYhat^{\Pgt(i)} \right)$
in the integrand.

The TF for the visible $\Pgt$ decay products and for the hadronic recoil are taken from Ref.~\cite{SVfitMEM}.
The resolution on the $\pT$ of $\tauh$ is modelled by the function:
\begin{equation}
W_{\Phadron}( \pT^{\vis} | \pThat^{\vis} ) = 
 \begin{cases}
   \mathcal{N} \, \xi_{1} \, \left( \frac{\alpha_{1}}{x_{1}} - x_{1} - \frac{x - \mu}{\sigma} \right)^{-\alpha_{1}} \,  
 & \mbox{if } x < x_{1} \\
   \mathcal{N} \, \exp\left( -\frac{1}{2} \, \left( \frac{x - \mu}{\sigma} \right)^{2} \right) \,
 & \mbox{if } x_{1} \leq x \leq x_{2} \\
   \mathcal{N} \, \xi_{2} \, \left( \frac{\alpha_{2}}{x_{2}} - x_{2} - \frac{x - \mu}{\sigma} \right)^{-\alpha_{2}} \, 
 & \mbox{if } x > x_{2} \, ,
 \end{cases}
\label{eq:tf_tauToHadDecays_pT}
\end{equation}
where we use the values $\mu = 1.0$, $\sigma = 0.03$, $x_{1} = 0.97$, $\alpha_{1} = 7$,
$x_{2} = 1.03$, and $\alpha_{2} = 3.5$ for its parameters,
while its $\eta$, $\phi$, and mass are assumed to be reconstructed with negligible experimental resolution.
The latter assumption is also made for the $\pT$, $\eta$, and $\phi$ of electrons and muons.
The momentum of the hadronic recoil is modelled by a two-dimensional normal distribution
and assumed to be reconstructed with a resolution of $\sigma_{\textrm{x}} = \sigma_{\textrm{y}} = 10$~\GeV 
on each of its components $\pX$ and $\pY$:
\begin{align}
W_{\rec}( \pX^{\rec},\pY^{\rec} | \pXhat^{\rec},\pYhat^{\rec} ) = & 
\frac{1}{2\pi \, \sqrt{\vert V \vert}} \, \exp \left( -\frac{1}{2}
  \left( \begin{array}{c} \Delta\pX^{\rec} \\ \Delta\pY^{\rec} \end{array} \right)^{T}
  \cdot V^{-1} \cdot
   \left( \begin{array}{c} \Delta\pX^{\rec} \\ \Delta\pY^{\rec} \end{array} \right)
  \right) \, , \nonumber \\
\quad \mbox{ with } \quad V = & \left( \begin{array}{cc} \sigma_{x}^{2} & 0 \\ 0 & \sigma_{y}^{2} \end{array} \right) \, .
\end{align}

The number of particles in the final state, $n$, depends on how many $\Pgt$ leptons decay to electrons or muons 
and how many decay to hadrons.
Following the formalism developed in Ref.~\cite{SVfitMEM}, 
we treat hadronic $\Pgt$ decays as two-body decays into a hadronic system $\tauh$ and a $\Pnut$.
Correspondingly, $n$ increases by $3$ for each $\Pgt$ lepton that decays to an electron or a muon
and by $2$ units for each $\Pgt$ lepton that decays hadronically.
Particles that are part of the hadronic recoil are treated as described in Section~2.2 of Ref.~\cite{SVfitMEM} 
and do not increase $n$.

\begin{figure}[h]
\begin{center}
\includegraphics*[height=58mm]{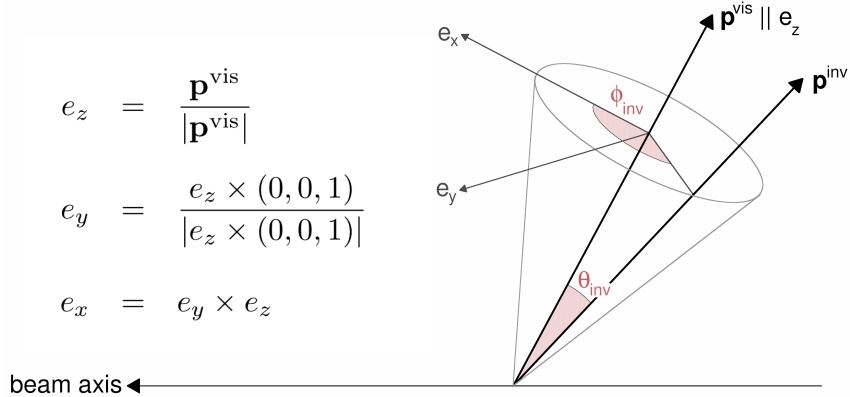}
\end{center}
\caption{
  Illustration of the variable $\phi_{\inv}$ that specifies the orientation of the $\bm{p}^{\inv}$ vector relative to the $\bm{p}^{\vis}$ vector.
  The angle $\theta_{\inv}$ between the $\bm{p}^{\inv}$ vector and the $\bm{p}^{\vis}$ vector is related to the variable $z$,
  as described in Section 2.4 of Ref.~\cite{SVfitMEM}, from which the illustration was taken.
}
\label{fig:tauDecayParametrization}
\end{figure}

The dimensionality of the integration over the phase-space element $d\Phi_{n}$ 
can be reduced by means of analytic transformations. 
Two (three) variables are sufficient to fully parametrize the kinematics of hadronic (leptonic) $\Pgt$ decays.
Following Ref.~\cite{SVfitMEM}, we choose to parametrize hadronic $\Pgt$ decays by the variables $z$ and $\phi_{\inv}$,
and leptonic $\Pgt$ decays by the variables $z$, $\phi_{\inv}$, and $m_{\inv}$.
The variable $z$ corresponds to the fraction of $\Pgt$ lepton energy, in the laboratory frame, that is carried by the visible $\Pgt$ decay products.
The variable $\phi_{\inv}$ specifies the orientation of the $\bm{p}^{\inv}$ vector relative to the $\bm{p}^{\vis}$ vector (see Fig.~\ref{fig:tauDecayParametrization} for illustration),
where the vector $\bm{p}^{\inv}$ refers to the vectorial sum of the momenta of the two neutrinos (to the momentum of the single $\Pnut$) produced in leptonic (hadronic) $\Pgt$ decays.
The variable $m_{\inv}$ denotes the mass of the neutrino pair produced in leptonic $\Pgt$ decays.

Expressions for the product of the phase-space element $d\Phi_{n}$ 
with the squared moduli $\vert \BW^{(i)}_{\Pgt} \vert^{2} \cdot \vert \mathcal{M}^{(i)}_{\Pgt\to\cdots}(\bm{\hat{p}}) \vert^{2}$ 
of the ME for the $\Pgt$ decays, 
obtained by the aforementioned transformations, are given by Eq.~(33) in Ref.~\cite{SVfitMEM}.
The expressions read:
\begin{align}
\vert \BW_{\Pgt} \vert^{2} \cdot \vert \mathcal{M}^{(i)}_{\Pgt\to\cdots}(\bm{\tilde{p}}) \vert^{2} \, d\Phi^{(i)}_{\tauhnu} 
 = & \, \frac{\pi}{m_{\Pgt} \, \Gamma_{\Pgt}} \,
 f_{\Phadron}\left(\bm{\hat{p}}^{\vis(i)}, m^{\vis(i)},
   \bm{\hat{p}}^{\inv(i)}\right) \, \frac{d^{3}\bm{\hat{p}}^{\vis}}{2 \hat{E}_{\vis}} \, dz \, d\phi_{\inv} \quad \mbox{ and } \nonumber \\
\vert \BW_{\Pgt} \vert^{2} \cdot \vert \mathcal{M}^{(i)}_{\Pgt\to\cdots}(\bm{\tilde{p}}) \vert^{2} \, d\Phi^{(i)}_{\ellnunu} 
 = & \, \frac{\pi}{m_{\Pgt} \, \Gamma_{\Pgt}} \, f_{\ell}\left(\bm{\hat{p}}^{\vis(i)},
 m^{\vis(i)}, \bm{\hat{p}}^{\inv(i)}\right) \, \frac{d^{3}\bm{\hat{p}}^{\vis}}{2 \hat{E}_{\vis}} \, dz \, dm^{2}_{\inv} \, d\phi_{\inv}
 \nonumber \, ,
\end{align}
where the functions $f_{\Phadron}$ and $f_{\ell}$ are defined as:
\begin{align}
f_{h}\left(\bm{p}^{\vis}, m_{\vis}, \bm{p}^{\inv}\right) = &
  \frac{\vert\mathcal{M}^{\eff}_{\Pgt \to \tauh\Pnut}\vert^{2}}{256\pi^{6}} \cdot \frac{E_{\vis}}{\vert\bm{p}^{\vis}\vert \, z^{2}} \quad \mbox{ and } \nonumber \\
f_{\Plepton}\left(\bm{p}^{\vis}, m_{\vis}, \bm{p}^{\inv}\right) = &
  \frac{I_{\inv}}{512\pi^{6}} \cdot \frac{E_{\vis}}{\vert\bm{p}^{\vis}\vert \, z^{2}} \nonumber \, , 
\end{align}
with:
\begin{align}
\vert \mathcal{M}^{\textrm{eff}}_{\Pgt \to \tauhnu} \vert^{2} = & 16 \pi \, m_{\Pgt} \, \Gamma_{\Pgt} \cdot
  \frac{m_{\Pgt}^{2}}{m_{\Pgt}^{2} - m_{\vis}^{2}} \cdot \mathcal{B}(\Pgt \to \textrm{hadrons} + \Pnut) \quad \mbox { and } \nonumber \\
I_{\inv} = & \, \frac{1}{2} \, m_{\inv} \, \int \, \frac{d\Omega_{v}}{(2\pi)^{3}} \, 
  \vert\mathcal{M}_{\Pgt \to \ellnunu}\vert^{2} \, , \quad \mbox { where } \nonumber \\ 
\vert\mathcal{M}_{\Pgt \to \ellnunu} \vert^{2} = & 64 \, G^{2}_{F} \,
  \left( E_{\Pgt} \, E_{\APnu_{\Plepton}} - \bm{p}^{\Pgt} \cdot
  \bm{p}^{\APnu_{\Plepton}} \right) \, \left( E_{\Plepton} \,
  E_{\Pnut} - \bm{p}^{\Plepton} \cdot \bm{p}^{\Pnut} \right) \nonumber 
\end{align}
and $\mathcal{B}(\Pgt \to \textrm{hadrons} + \Pnut) = 0.648$~\cite{PDG} 
denotes the measured branching fraction for $\Pgt$ leptons to decay hadronically.

The knowledge that the four $\Pgt$ leptons originate from the decay of two $\PHiggs$ bosons is incorporated into the likelihood function 
$\mathcal{P}$ by suitably chosen constraints.
For the purpose of defining the constraints, it is useful to enumerate the $\Pgt$ leptons 
such that the two $\Pgt$ leptons with indices $i=1$ and $i=2$ (and similarly the two $\Pgt$ leptons with indices $i=3$ and $i=4$) are interpreted as originating from the same $\PHiggs$ boson.
We then require that the visible $\Pgt$ decay products corresponding to the indices $i=1$ and $i=2$ have opposite charge,
and the same applies to the visible $\Pgt$ decay products corresponding to the indices $i=3$ and $i=4$.
As the width of the $\PHiggs$ boson is known to be small~\cite{HIG-14-002,Aad:2015xua} compared to the experimental resolution that we aim to achieve on $m_{\PHiggs\PHiggs}$,
we choose to neglect it and use the narrow-width approximation (NWA) for each $\PHiggs$ boson.
The NWA introduces two $\delta$-functions, 
\begin{align}
 & \delta\left( (\hat{E}_{\Pgt(1)} + \hat{E}_{\Pgt(2)})^{2} - (\bm{\hat{p}}^{\Pgt(1)} + \bm{\hat{p}}^{\Pgt(2)})^{2} - m_{\PHiggs}^{2} \right) \quad \mbox{ and } \nonumber \\
 & \delta\left( (\hat{E}_{\Pgt(3)} + \hat{E}_{\Pgt(4)})^{2} - (\bm{\hat{p}}^{\Pgt(3)} + \bm{\hat{p}}^{\Pgt(4)})^{2} - m_{\PHiggs}^{2} \right)
\end{align}
into Eq.~(\ref{eq:likelihood}).
For the purpose of evaluating the $\delta$-functions,
we make the simplifying assumption that the angle between the vectors $\bm{\hat{p}}^{\vis(i)}$ and $\bm{\hat{p}}^{\inv(i)}$ is negligible.
The assumption is justified by the fact that at the LHC the $\pT$ of the visible $\Pgt$ decay products are typically large compared to the mass, 
$m_{\Pgt} = 1.777$~\GeV~\cite{PDG}, of the $\Pgt$ lepton.
With this assumption, the $\delta$-functions simplify to:
\begin{equation*}
\delta\left(\frac{m_{\vis(12)}}{z_{1} \, z_{2}} - m_{\PHiggs}^{2}\right) \quad \mbox{ and } \quad \delta\left(\frac{m_{\vis(34)}}{z_{3} \, z_{4}} - m_{\PHiggs}^{2}\right) \, ,
\end{equation*}
where we denote by the symbol $m_{\vis(ij)}$ the ``visible mass'' of the decay products of $\Pgt$ leptons $i$ and $j$:
\begin{equation*}
m_{\vis(ij)} = (\hat{E}_{\vis(i)} + \hat{E}_{\vis(i)})^{2} - (\bm{\hat{p}}^{\vis(i)} + \bm{\hat{p}}^{\vis(j)})^{2} \, .
\end{equation*}
The $\delta$-functions are used to eliminate the integration over the variables $z_{2}$ and $z_{4}$.
The $\delta$-function rule,
\begin{equation*} 
\delta \left( g(x) \right) = \sum_{k} \, \frac{\delta \left( x - x_{k}
  \right)}{\vert g'(x_{k}) \vert} \, ,
\end{equation*}
where the sum extends over all roots $x_{k}$ of the function $g(x)$,
yields the two factors:
\begin{equation*}
\frac{z_{2}}{m_{\PHiggs}^{2}} \quad \mbox{ and } \quad \frac{z_{4}}{m_{\PHiggs}^{2}} \, ,
\end{equation*}
with the roots:
\begin{equation*}
z_{2} = \frac{m_{\vis(12)}}{m_{\PHiggs}^{2} \, z_{1}} \quad \mbox{ and } \quad z_{4} = \frac{m_{\vis(34)}}{m_{\PHiggs}^{2} \, z_{3}} \, .
\end{equation*}

The condition
$\delta\left( \left(\sum_{i=1}^{4} \, E_{\Pgt(i)}\right)^{2} - \left(\sum_{i=1}^{4} \, \bm{\hat{p}}^{\Pgt(i)}\right)^{2} - m_{\textrm{X}} \right)$
is used to eliminate the integration over the variable $z_{3}$.
It yields the factor:
\begin{equation}
\lvert \frac{z_{1} \, z_{3}^{2}}{b \, z_{3}^{2} - c} \rvert \, ,
\label{eq:deltaFuncFactor}
\end{equation}
with the two roots:
\begin{equation*}
z_{3}^{(+)} = \frac{a + \sqrt{b}}{c} \quad \mbox{ and } \quad z_{3}^{(-)} = \frac{a - \sqrt{b}}{c} \, ,
\end{equation*}
where:
\begin{align}
a = & (m_{\textrm{X}}^{2} - 2 \, m_{\PHiggs}^{2}) \, z_{1} \, , \nonumber \\
b = & \frac{m_{\vis(14)}^{2}}{m_{\vis(34)}^{2}} \, m_{\PHiggs}^{2} + \frac{m_{\vis(24)}^{2}}{m_{\vis(12)}^{2} \, m_{\vis(34)}^{2}} \, m_{\PHiggs}^{4} \, z_{1}^{2} \quad \mbox{ and } \nonumber \\
c = & m_{\vis(13)}^{2} + \frac{m_{\vis(23)}^{2}}{m_{\vis(12)}^{2}} \, z_{1}^{2} \, .
\end{align}
The requirement that the energies of electrons, muons, and $\tauh$ 
as well as the energies of the neutrinos produced in the $\Pgt$ decays are positive
restricts the variable $z_{3}$ to the range $0 < z_{3} \leq 1$.
In case the roots $z_{3}^{(+)}$ and $z_{3}^{(-)}$ are both within this range,
the integrand is evaluated for each root separately and the values obtained for each root are summed.
Otherwise, only the root satisfying the condition $0 < z_{3} \leq 1$ is retained.

Expressions for the likelihood function $\mathcal{P}$, obtained after performing these analytic transformations, 
are given by Eqs.~(\ref{eq:likelihood_thththth}) to~(\ref{eq:likelihood_llll}) in the Appendix.
We refer to the different decay channels of the four $\Pgt$ leptons as 
$\tauh\tauh\tauh\tauh$, $\Plepton\tauh\tauh\tauh$, $\Plepton\Plepton\tauh\tauh$, $\Plepton\Plepton\Plepton\tauh$, and $\Plepton\Plepton\Plepton\Plepton$, 
where the symbol $\Plepton$ refers to an electron or muon,
and the neutrinos produced in the $\Pgt$ decays are omitted from the nomenclature.
The dimension of integration varies between $5$ for events in the $\tauh\tauh\tauh\tauh$ decay channel and $9$ for events in the $\Plepton\Plepton\Plepton\Plepton$ channel.
The expressions given in the Appendix correspond to one particular association of reconstructed electrons, muons, and $\tauh$ 
to the indices $1$, $2$, $3$, and $4$, which enumerate the $\Pgt$ decay products in Eqs.~(\ref{eq:likelihood_thththth}) to~(\ref{eq:likelihood_llll}).
Expressions for alternative associations can be obtained by appropriate permutations of the indices.

For any one of these associations
the best estimate, $m_{\PHiggs\PHiggs}$, for the mass of the $\PHiggs$ boson pair is obtained 
by finding the value of $m_{\textrm{X}}$ that maximizes the value of $\mathcal{P}$.
The integrand in Eqs.~(\ref{eq:likelihood_thththth}) to~(\ref{eq:likelihood_llll})
is evaluated for a series of mass hypotheses $m_{\textrm{X}}^{(i)}$.
Starting from the initial value $m_{\textrm{X}}^{(0)} = 1.0125 \cdot \max (2 \, m_{\PHiggs}, m_{\PHiggs\PHiggs}^{\vis})$,
where
$m_{\PHiggs\PHiggs}^{\vis} = \sqrt{\left(\sum_{i=1}^{4} \, E_{\vis(i)}\right)^{2} - \left(\sum_{i=1}^{4} \, \bm{p}^{\vis(i)}\right)^{2}}$,
the next mass hypothesis in the series is defined by the recursive relation $m_{\textrm{X}}^{(i+1)} = (1 + \delta) \cdot m_{\textrm{X}}^{(i)}$.
The step size $\delta = 0.025$ is chosen such that it is small compared to the resolution on $m_{\PHiggs\PHiggs}$
that we expect our algorithm to achieve.
The evaluation of the integral is performed numerically, using the VAMP algorithm~\cite{VAMP},
an improved implementation of the VEGAS algorithm~\cite{VEGAS}.
For each mass hypothesis $m_{\textrm{X}}^{(i)}$, the integrand is evaluated $20\,000$ times.

We note in passing that our algorithm alternatively supports an integration methods based on a custom implementation
of the Markov-Chain integration method with the Metropolis--Hastings algorithm~\cite{Metropolis_Hastings}.
The latter allows to reconstruct the $\pT$, pseudo-rapidity $\eta$, and azimuthal angle $\phi$ of the resonance $\textrm{X}$ also.
In this paper, we focus on the reconstruction of the mass, however.

A remaining issue for the algorithm is that in $\PHiggs\PHiggs \to \Pgt\Pgt\Pgt\Pgt$ events 
there exist two possibilities for building pairs of $\Pgt$ leptons of opposite charge.
The ambiguity is resolved, and a unique value of $m_{\PHiggs\PHiggs}$ is obtained for each event, 
by first discarding pairings for which either $m_{\vis(12)}$ or $m_{\vis(34)}$ exceeds $m_{\PHiggs}$
and then selecting the pairing for which the likelihood function $\mathcal{P}$
attains the maximal value (for any $m_{\textrm{X}}$).
We will demonstrate in Section~\ref{sec:performance} that this choice yields the correct pairing for the majority of events.

%% file: performance.tex
\section{Performance}
\label{sec:performance}

The performance of the algorithm is quantified in terms of the resolution achieved in reconstructing $m_{\PHiggs\PHiggs}$.
The resolution is studied using simulated samples of events
in which a heavy resonance $\textrm{X}$ decays into a pair of $\PHiggs$ bosons,
and the $\PHiggs$ bosons subsequently decay to four $\Pgt$ leptons.
Samples are produced for $m_{\textrm{X}} = 300$, $500$, and $800$~\GeV.
We expect the resolution to be similar for resonances of spin $0$ and spin $2$,
but focus on studying resonances of spin $0$ in this paper.
Events are generated for proton-proton collisions at $\sqrt{s} = 13$~\TeV centre-of-mass energy,
using the leading order program MadGraph, in the version MadGraph\_aMCatNLO 2.3.2.2~\cite{MadGraph_aMCatNLO},
with the NNPDF3.0 set of parton distribution functions~\cite{NNPDF1,NNPDF2,NNPDF3}.
Parton shower and hadronization processes are modelled using the generator PYTHIA 8.2~\cite{pythia8} with the tune CUETP8M1~\cite{PYTHIA_CUETP8M1tune_CMS}.
The decays of $\Pgt$ leptons, including polarization effects, are modelled by PYTHIA.

We select events in the decay channel $\textrm{X} \to \PHiggs\PHiggs \to \Pgt\Pgt\Pgt\Pgt \to \Plepton\Plepton\tauh\tauh$
and study them on generator level.
Reconstruction effects are simulated by varying the generator-level quantities within their experimental resolution,
which we perform by randomly sampling from the TF 
$W_{\Phadron}( \pT^{\vis} | \pThat^{\vis} )$
and
$W_{\rec}( \pX^{\rec},\pY^{\rec} | \pXhat^{\rec},\pYhat^{\rec} )$
described in Section~\ref{sec:algorithm}.
The electrons, muons, and $\tauh$ are required to satisfy conditions on $\pT$ and $\eta$, 
which are typical for data analyses performed by the ATLAS and CMS collaborations during LHC Run $2$.
Electrons (muons) are required to be within $\vert\eta\vert < 2.5$ ($\vert\eta\vert < 2.4$).
The lepton of higher (lower) $\pT$ is required to pass a $\pT$ threshold of $25$ ($15$)~\GeV.
Each of the two $\tauh$ is required to satisfy $\pT > 20$~\GeV and $\vert\eta\vert < 2.3$.

\begin{figure}
\setlength{\unitlength}{1mm}
\begin{center}
\begin{picture}(180,212)(0,0)
\put(-4.5, 152.0){\mbox{\includegraphics*[height=60mm]
  {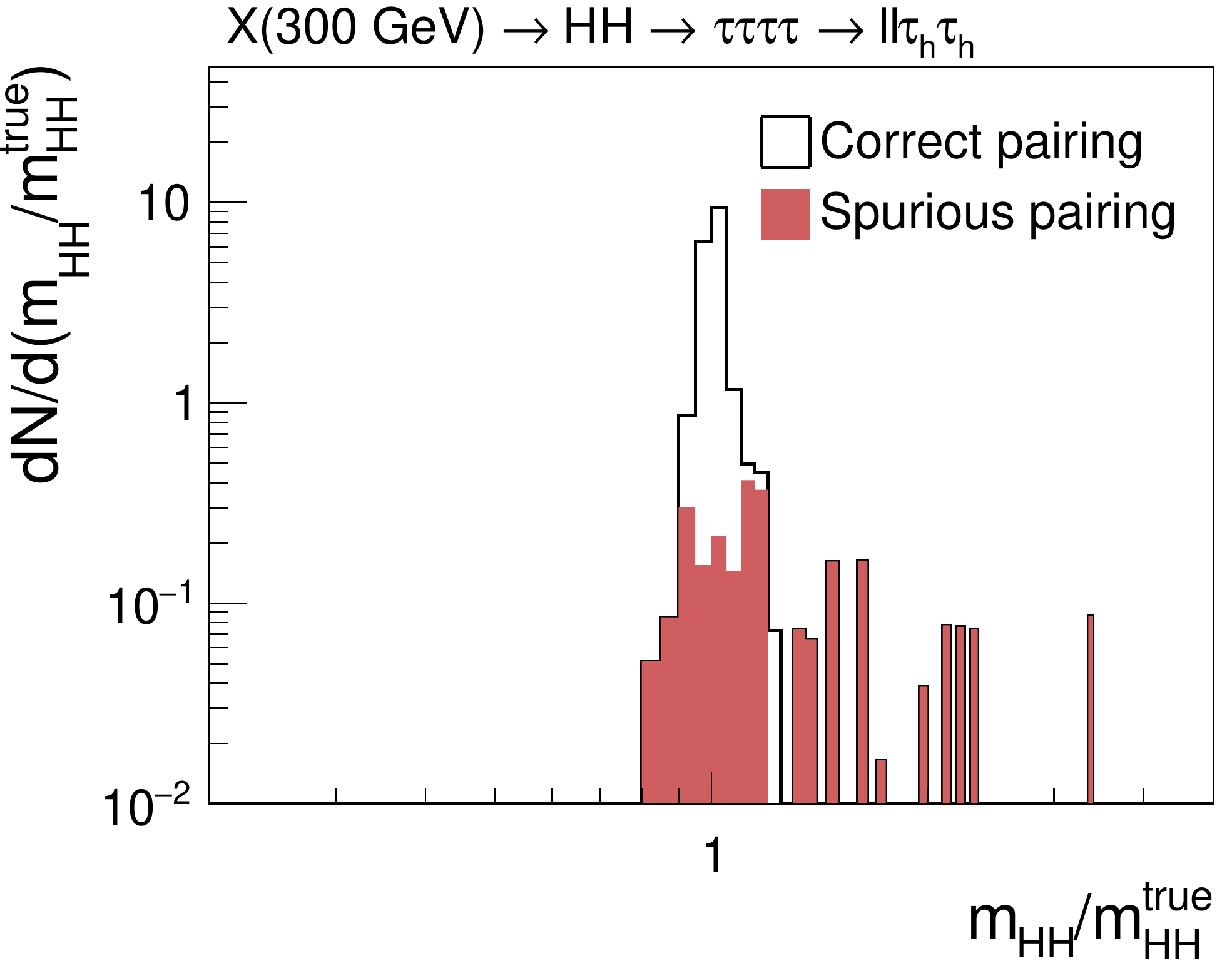}}}
\put(81.5, 152.0){\mbox{\includegraphics*[height=60mm]
  {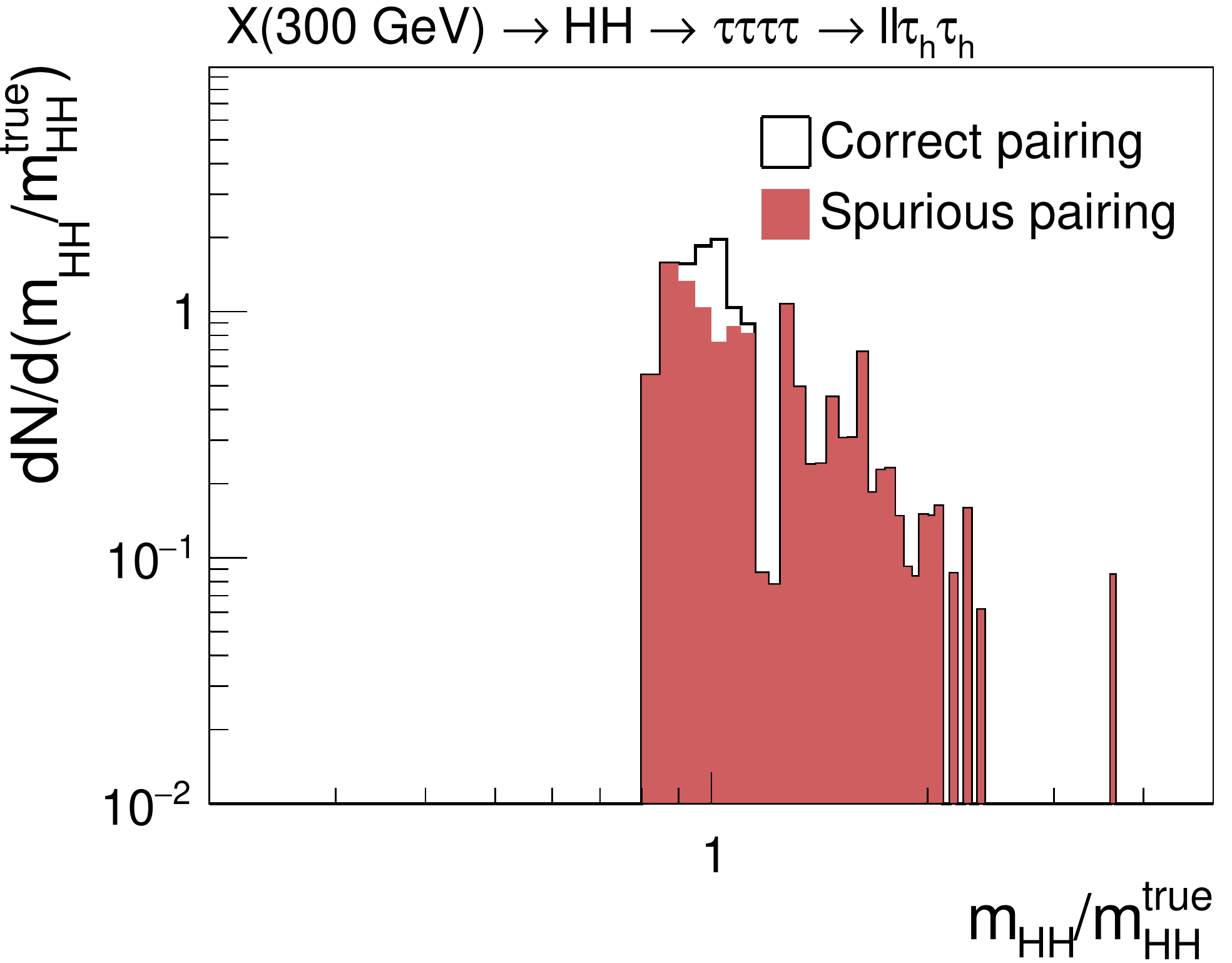}}}
\put(-4.5, 78.0){\mbox{\includegraphics*[height=60mm]
  {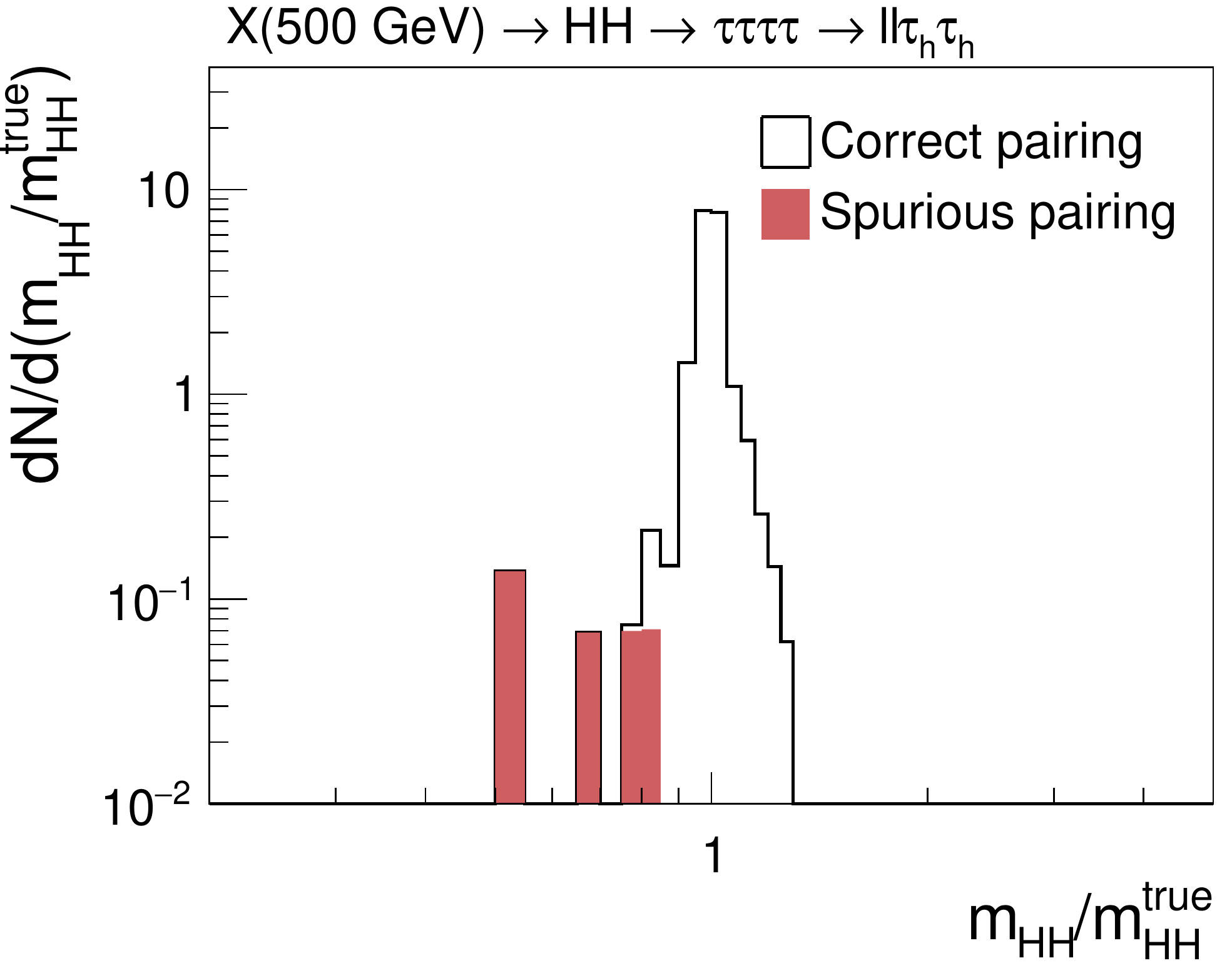}}}
\put(81.5, 78.0){\mbox{\includegraphics*[height=60mm]
  {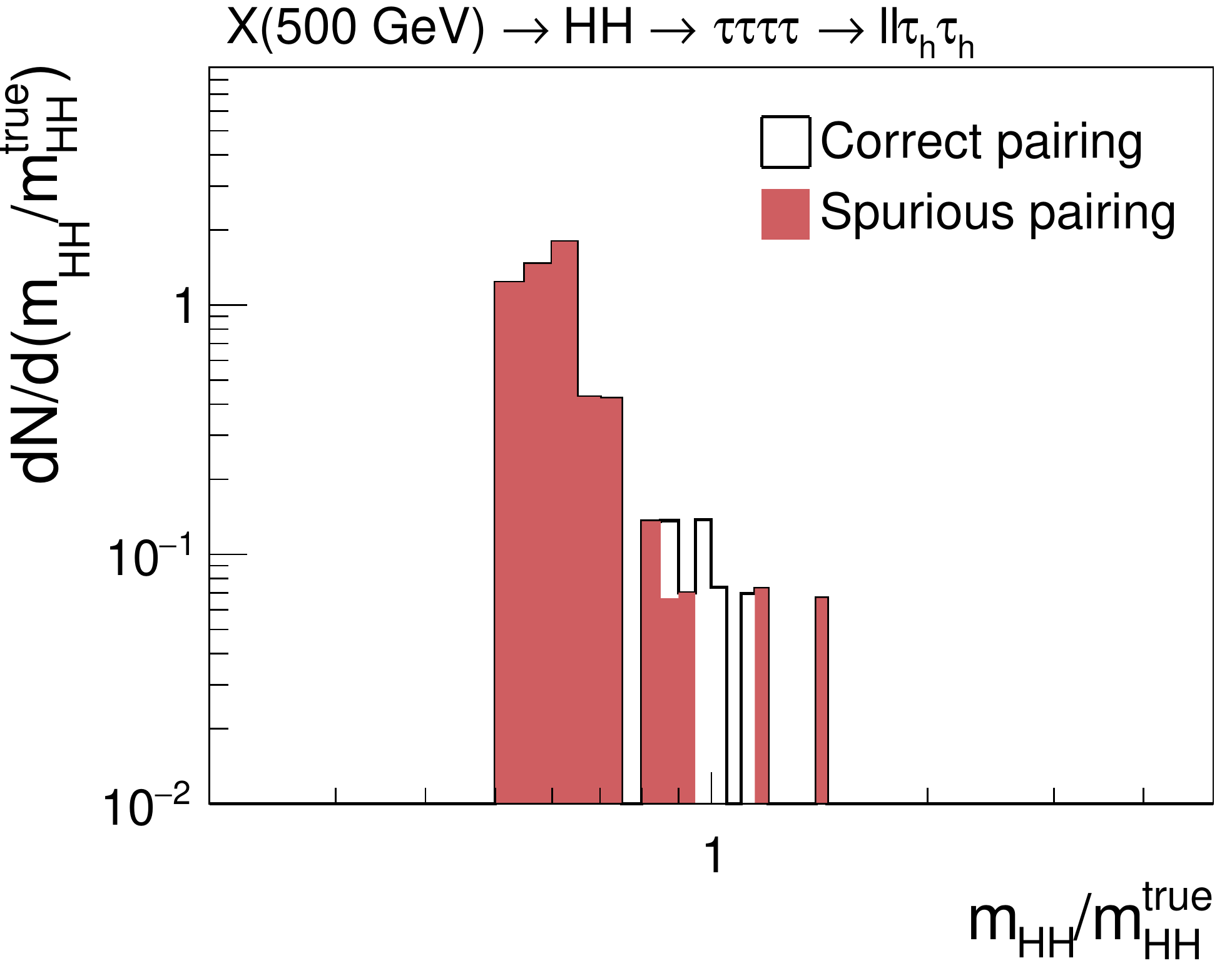}}}
\put(-4.5, 4.0){\mbox{\includegraphics*[height=60mm]
  {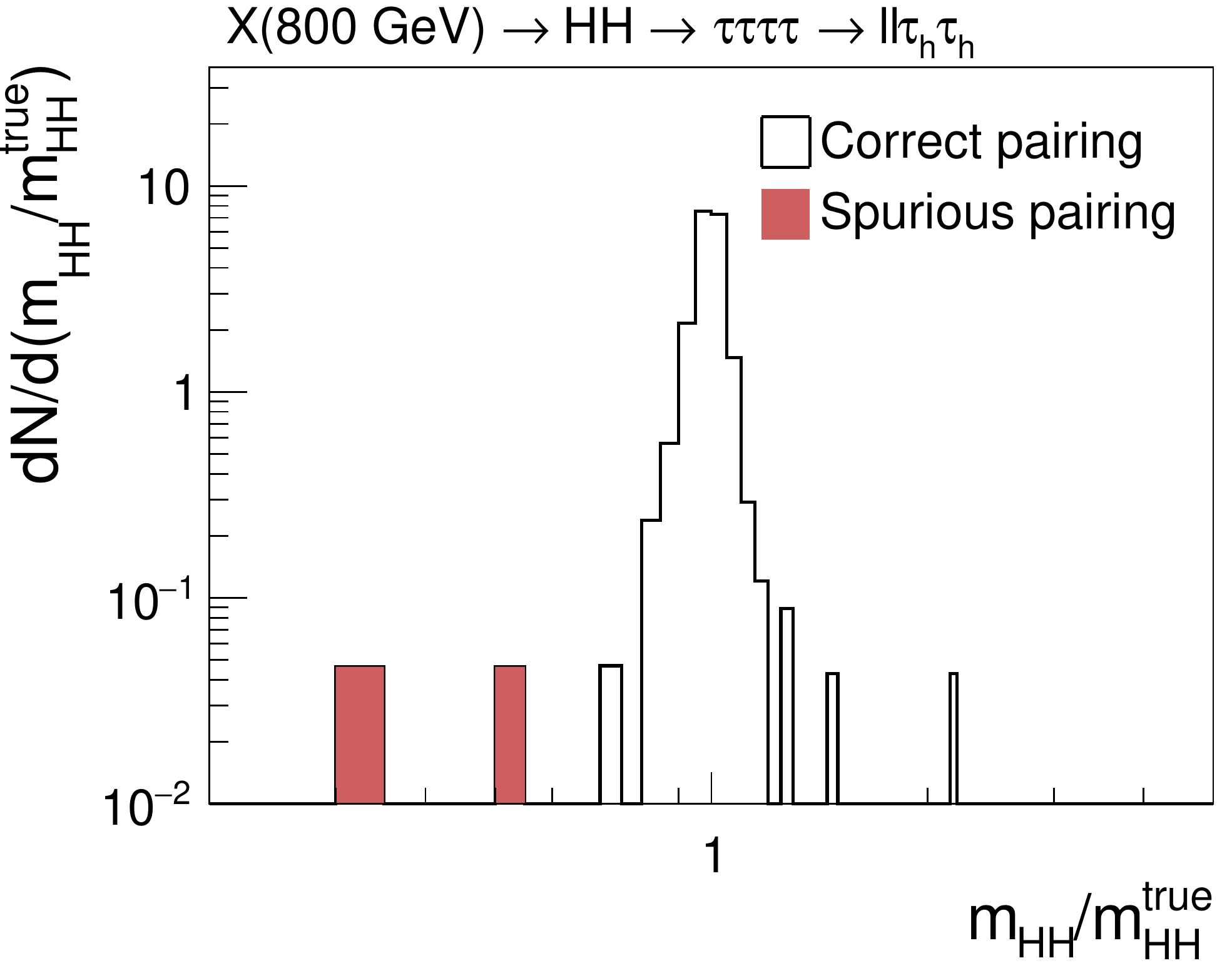}}}
\put(81.5, 4.0){\mbox{\includegraphics*[height=60mm]
  {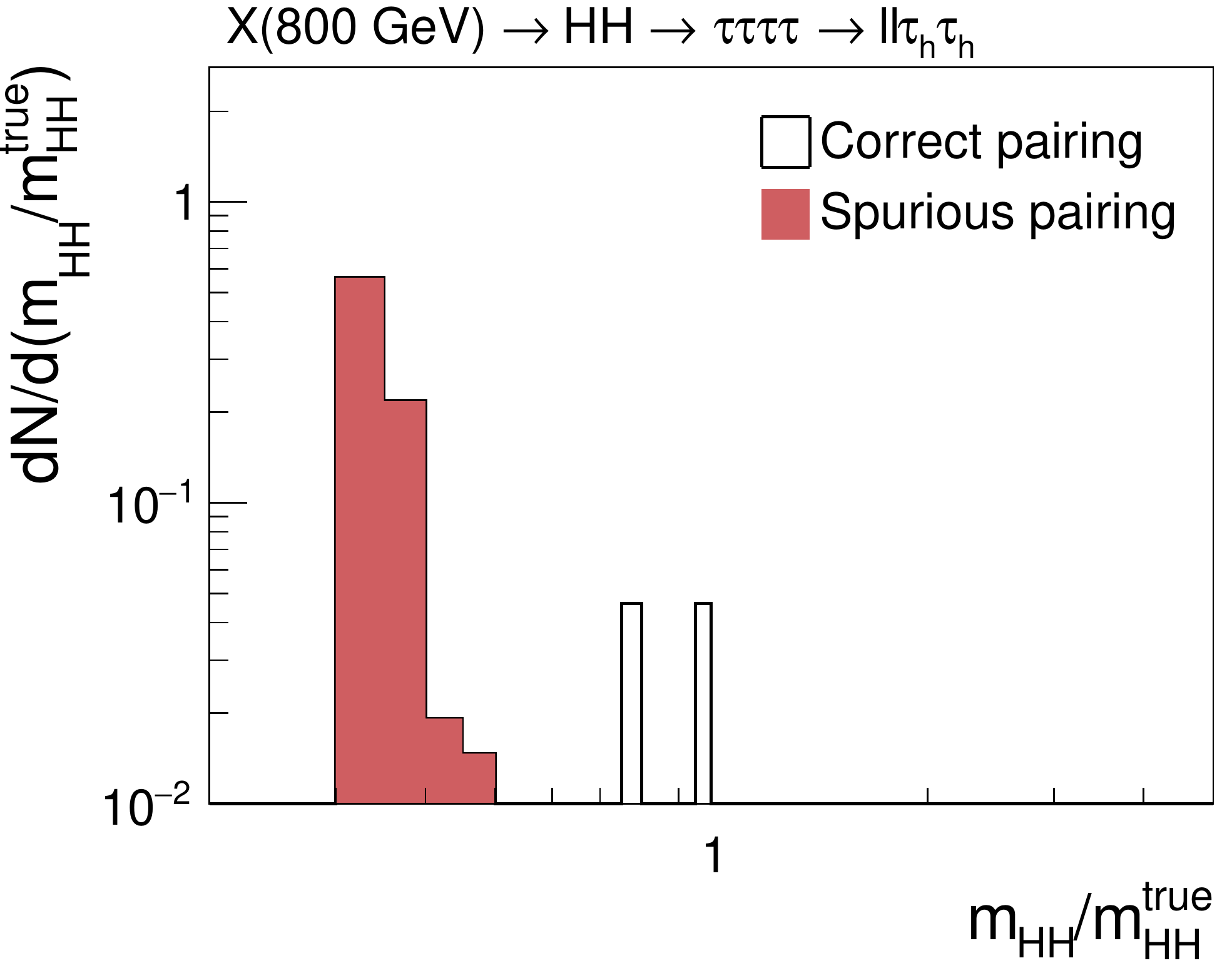}}}
\put(38.0, 148.0){\small (a)}
\put(124.0, 148.0){\small (b)}
\put(38.0, 74.0){\small (c)}
\put(124.0, 74.0){\small (d)}
\put(38.0, 0.0){\small (e)}
\put(124.0, 0.0){\small (f)}
\end{picture}
\end{center}
\caption{
  Distributions in $m_{\PHiggs\PHiggs}$, relative to the true mass $m_{\PHiggs\PHiggs}^{\textrm{true}} = m_{\textrm{X}}$ of the $\PHiggs$ boson pair,
  in events in which a heavy resonance $\textrm{X}$ of mass $m_{\textrm{X}} = 300$ (a,b), $500$ (c,d), and $800$~\GeV (e,f)
  decays via $\textrm{X} \to \PHiggs\PHiggs \to \Pgt\Pgt\Pgt\Pgt \to \Plepton\Plepton\tauh\tauh$.
  The distributions are shown separately for the chosen (a,c,e) and for the discarded (b,d,f) pairings,
  and are further subdivided into correct and spurious pairings.
  The axis of abscissae ranges from $0.2$ to $5$.
}
\label{fig:massDistributions}
\end{figure}

The resolution on $m_{\PHiggs\PHiggs}$ is studied in terms of the ratio between the reconstructed value of $m_{\PHiggs\PHiggs}$ 
and the true mass $m_{\PHiggs\PHiggs}^{\textrm{true}}$ of the $\PHiggs$ boson pair.
Distributions in this ratio are shown in Fig.~\ref{fig:massDistributions}.
They are shown separately for chosen 
(pairings that maximize the likelihood function $\mathcal{P}$) 
and for discarded (other) pairings and for
events in which electrons, muons, and $\tauh$ are correctly associated to $\PHiggs$ boson pairs and events with spurious pairings.
The correct pairing is chosen in $87$, $98$, and $>99\%$ of the events with $m_{\textrm{X}} = 300$, $500$, and $800$~\GeV, respectively.
The resolution on $m_{\PHiggs\PHiggs}$ for the chosen pairings amounts to $22$, $7$, and $9\%$,
relative to the true mass of the $\PHiggs$ boson pair.
The mass resolution for resonances of $m_{\textrm{X}} = 300$, near the kinematic threshold $m_{\textrm{X}} \approx 2 \, m_{\PHiggs}$,
is limited by the fact that the wrong pairing is chosen in $13\%$ of events.
For events in which the correct pairing is chosen, the resolution on $m_{\PHiggs\PHiggs}$ amounts to $4$, $6$, and $8\%$
for $m_{\textrm{X}} = 300$, $500$, and $800$~\GeV, respectively.
We leave the optimization of the choice of the correct pairing for resonances of low mass to future studies.

The algorithm requires typically $2$s of CPU time per event to reconstruct $m_{\PHiggs\PHiggs}$.

%% file: summary.tex
\section{Summary}
\label{sec:summary}

An algorithm for the reconstruction of the mass $m_{\PHiggs\PHiggs}$ of the $\PHiggs$ boson pair 
in events in which the Higgs boson pair decays via $\PHiggs\PHiggs \to \Pgt\Pgt\Pgt\Pgt$
and the $\Pgt$ leptons subsequently decay into electrons, muons, or hadrons has been presented.
The resolution on $m_{\PHiggs\PHiggs}$ has been studied in simulated events 
and amounts to $22$, $7$, and $9\%$, relative to the true mass of the $\PHiggs$ boson pair,
for events containing resonances $\textrm{X}$ of mass $m_{\textrm{X}} = 300$, $500$, and $800$~\GeV, respectively.
The mass resolution for resonances of low mass, near the kinematic threshold $m_{\textrm{X}} \approx 2 \, m_{\PHiggs}$,
is limited by the fact that the algorithm chooses the wrong association of electrons, muons, and $\tauh$ 
to $\PHiggs$ boson pairs in $13\%$ of events.
The probability to choose a spurious pairing decreases for resonances of higher mass and becomes negligible for $m_{\textrm{X}} \gtrsim 500$~\GeV.
The optimization of the choice of the correct pairing for resonances of low mass is left to future studies.
We expect that our algorithm will be useful in searches for heavy resonances decaying to $\PHiggs$ boson pairs
at the LHC.

%% file: appendix.tex
\section{Appendix}
\label{sec:appendix}

\subsubsection{$\PHiggs\PHiggs \to \Pgt\Pgt\Pgt\Pgt \to \tauh\tauh\tauh\tauh$ decay channel}

\begin{align}
&
\mathcal{P}(\bm{p}^{\vis(1)},\bm{p}^{\vis(2)},\bm{p}^{\vis(3)},\bm{p}^{\vis(4)};\pX^{\rec},\pY^{\rec}|m_{\textrm{X}})
= \frac{32\pi^{8}}{m_{\Pgt} \, \Gamma_{\Pgt} \, s} \, \nonumber \\
& \qquad \int \, dz_{1} \, d\phi_{\inv(1)} \, d\phi_{\inv(2)} \, d\phi_{\inv(3)} \, d\phi_{\inv(4)} \, 
  \sum_{z_{3}^{+},z_{3}^{-}} \, \Bigr\lvert \frac{z_{1} \, z_{3}^{2}}{b \, z_{3}^{2} - c} \Bigr\rvert \cdot \nonumber \\
& \qquad \frac{\vert\mathcal{M}^{\eff(1)}_{\Pgt \to \tauh\Pnut}\vert^{2}}{256\pi^{6}} 
  \frac{E_{\vis(1)}}{\vert\bm{p}^{\vis(1)}\vert \, z_{1}^{2}} \, \frac{d\pThat^{\vis(1)}}{2 \, \hat{E}_{\vis(1)}} 
  \cdot W_{\Phadron}( \pT^{\vis(1)} | \pThat^{\vis(1)} ) \cdot \nonumber \\
& \qquad \frac{\vert\mathcal{M}^{\eff(2)}_{\Pgt \to \tauh\Pnut}\vert^{2}}{256\pi^{6}} 
  \frac{E_{\vis(2)}}{\vert\bm{p}^{\vis(2)}\vert \, z_{2}^{2}} \, \frac{d\pThat^{\vis(2)}}{2 \, \hat{E}_{\vis(2)}} 
  \cdot W_{\Phadron}( \pT^{\vis(2)} | \pThat^{\vis(2)} ) \cdot \nonumber \\
& \qquad \frac{\vert\mathcal{M}^{\eff(3)}_{\Pgt \to \tauh\Pnut}\vert^{2}}{256\pi^{6}} 
  \frac{E_{\vis(3)}}{\vert\bm{p}^{\vis(3)}\vert \, z_{3}^{2}} \, \frac{d\pThat^{\vis(3)}}{2 \, \hat{E}_{\vis(3)}} 
  \cdot W_{\Phadron}( \pT^{\vis(3)} | \pThat^{\vis(3)} ) \cdot \nonumber \\
& \qquad \frac{\vert\mathcal{M}^{\eff(4)}_{\Pgt \to \tauh\Pnut}\vert^{2}}{256\pi^{6}} 
  \frac{E_{\vis(4)}}{\vert\bm{p}^{\vis(4)}\vert \, z_{4}^{2}} \, \frac{d\pThat^{\vis(4)}}{2 \, \hat{E}_{\vis(4)}} 
  \cdot W_{\Phadron}( \pT^{\vis(4)} | \pThat^{\vis(4)} ) \cdot \nonumber \\
& \qquad W_{\rec}( \pX^{\rec},\pY^{\rec} | \pXhat^{\rec},\pYhat^{\rec} ) 
\label{eq:likelihood_thththth}
\end{align}

\subsubsection{$\PHiggs\PHiggs \to \Pgt\Pgt\Pgt\Pgt \to \Plepton\tauh\tauh\tauh$ decay channel}

\begin{align}
&
\mathcal{P}(\bm{p}^{\vis(1)},\bm{p}^{\vis(2)},\bm{p}^{\vis(3)},\bm{p}^{\vis(4)};\pX^{\rec},\pY^{\rec}|m_{\textrm{X}})
= \frac{32\pi^{8}}{m_{\Pgt} \, \Gamma_{\Pgt} \, s} \, \nonumber \\
& \qquad \int \, dz_{1} \, dm^{2}_{\inv(1)} \, d\phi_{\inv(1)} \, d\phi_{\inv(2)} \, d\phi_{\inv(3)} \, d\phi_{\inv(4)} \, 
  \sum_{z_{3}^{+},z_{3}^{-}} \, \Bigr\lvert \frac{z_{1} \, z_{3}^{2}}{b \, z_{3}^{2} - c} \Bigr\rvert \cdot \nonumber \\
& \qquad \frac{I_{\inv(1)}}{512\pi^{6}} 
  \frac{E_{\vis(1)}}{\vert\bm{p}^{\vis(1)}\vert \, z_{1}^{2}} \, \frac{1}{2 \, \hat{E}_{\vis(1)}} 
  \cdot \frac{\vert\mathcal{M}^{\eff(2)}_{\Pgt \to \tauh\Pnut}\vert^{2}}{256\pi^{6}} 
  \frac{E_{\vis(2)}}{\vert\bm{p}^{\vis(2)}\vert \, z_{2}^{2}} \, \frac{d\pThat^{\vis(2)}}{2 \, \hat{E}_{\vis(2)}} 
  \cdot W_{\Phadron}( \pT^{\vis(2)} | \pThat^{\vis(2)} ) \cdot \nonumber \\
& \qquad \frac{\vert\mathcal{M}^{\eff(3)}_{\Pgt \to \tauh\Pnut}\vert^{2}}{256\pi^{6}} 
  \frac{E_{\vis(3)}}{\vert\bm{p}^{\vis(3)}\vert \, z_{3}^{2}} \, \frac{d\pThat^{\vis(3)}}{2 \, \hat{E}_{\vis(3)}} 
  \cdot W_{\Phadron}( \pT^{\vis(3)} | \pThat^{\vis(3)} ) \cdot \nonumber \\
& \qquad \frac{\vert\mathcal{M}^{\eff(4)}_{\Pgt \to \tauh\Pnut}\vert^{2}}{256\pi^{6}} 
  \frac{E_{\vis(4)}}{\vert\bm{p}^{\vis(4)}\vert \, z_{4}^{2}} \, \frac{d\pThat^{\vis(4)}}{2 \, \hat{E}_{\vis(4)}} 
  \cdot W_{\Phadron}( \pT^{\vis(4)} | \pThat^{\vis(4)} ) \cdot \nonumber \\
& \qquad W_{\rec}( \pX^{\rec},\pY^{\rec} | \pXhat^{\rec},\pYhat^{\rec} ) 
\label{eq:likelihood_lththth}
\end{align}

\subsubsection{$\PHiggs\PHiggs \to \Pgt\Pgt\Pgt\Pgt \to \Plepton\Plepton\tauh\tauh$ decay channel}

\begin{align}
&
\mathcal{P}(\bm{p}^{\vis(1)},\bm{p}^{\vis(2)},\bm{p}^{\vis(3)},\bm{p}^{\vis(4)};\pX^{\rec},\pY^{\rec}|m_{\textrm{X}})
= \frac{32\pi^{8}}{m_{\Pgt} \, \Gamma_{\Pgt} \, s} \, \nonumber \\
& \qquad \int \, dz_{1} \, dm^{2}_{\inv(1)} \, d\phi_{\inv(1)} \, dm^{2}_{\inv(2)}\, d\phi_{\inv(2)} \, d\phi_{\inv(3)} \, d\phi_{\inv(4)} \, 
  \sum_{z_{3}^{+},z_{3}^{-}} \, \Bigr\lvert \frac{z_{1} \, z_{3}^{2}}{b \, z_{3}^{2} - c} \Bigr\rvert \cdot \nonumber \\
& \qquad \frac{I_{\inv(1)}}{512\pi^{6}} 
  \frac{E_{\vis(1)}}{\vert\bm{p}^{\vis(1)}\vert \, z_{1}^{2}} \, \frac{1}{2 \, \hat{E}_{\vis(1)}}
  \cdot \frac{I_{\inv(2)}}{512\pi^{6}} 
  \frac{E_{\vis(2)}}{\vert\bm{p}^{\vis(2)}\vert \, z_{2}^{2}} \, \frac{1}{2 \, \hat{E}_{\vis(2)}} \cdot \nonumber \\
& \qquad \frac{\vert\mathcal{M}^{\eff(3)}_{\Pgt \to \tauh\Pnut}\vert^{2}}{256\pi^{6}} 
  \frac{E_{\vis(3)}}{\vert\bm{p}^{\vis(3)}\vert \, z_{3}^{2}} \, \frac{d\pThat^{\vis(3)}}{2 \, \hat{E}_{\vis(3)}} 
  \cdot W_{\Phadron}( \pT^{\vis(3)} | \pThat^{\vis(3)} ) \cdot \nonumber \\
& \qquad \frac{\vert\mathcal{M}^{\eff(4)}_{\Pgt \to \tauh\Pnut}\vert^{2}}{256\pi^{6}} 
  \frac{E_{\vis(4)}}{\vert\bm{p}^{\vis(4)}\vert \, z_{4}^{2}} \, \frac{d\pThat^{\vis(4)}}{2 \, \hat{E}_{\vis(4)}} 
  \cdot W_{\Phadron}( \pT^{\vis(4)} | \pThat^{\vis(4)} ) \cdot \nonumber \\
& \qquad W_{\rec}( \pX^{\rec},\pY^{\rec} | \pXhat^{\rec},\pYhat^{\rec} ) 
\label{eq:likelihood_llthth}
\end{align}

\subsubsection{$\PHiggs\PHiggs \to \Pgt\Pgt\Pgt\Pgt \to \Plepton\Plepton\Plepton\tauh$ decay channel}

\begin{align}
&
\mathcal{P}(\bm{p}^{\vis(1)},\bm{p}^{\vis(2)},\bm{p}^{\vis(3)},\bm{p}^{\vis(4)};\pX^{\rec},\pY^{\rec}|m_{\textrm{X}})
= \frac{32\pi^{8}}{m_{\Pgt} \, \Gamma_{\Pgt} \, s} \, \nonumber \\
& \qquad \int \, dz_{1} \, dm^{2}_{\inv(1)} \, d\phi_{\inv(1)} \, dm^{2}_{\inv(2)} \, d\phi_{\inv(2)} \, dm^{2}_{\inv(3)} \, d\phi_{\inv(3)} \, d\phi_{\inv(4)} \, 
  \sum_{z_{3}^{+},z_{3}^{-}} \, \Bigr\lvert \frac{z_{1} \, z_{3}^{2}}{b \, z_{3}^{2} - c} \Bigr\rvert \cdot \nonumber \\
& \qquad \frac{I_{\inv(1)}}{512\pi^{6}} 
  \frac{E_{\vis(1)}}{\vert\bm{p}^{\vis(1)}\vert \, z_{1}^{2}} \, \frac{1}{2 \, \hat{E}_{\vis(1)}} 
  \cdot \frac{I_{\inv(2)}}{512\pi^{6}} 
  \frac{E_{\vis(2)}}{\vert\bm{p}^{\vis(2)}\vert \, z_{2}^{2}} \, \frac{1}{2 \, \hat{E}_{\vis(2)}} \cdot \nonumber \\
& \qquad \frac{I_{\inv(3)}}{512\pi^{6}} 
  \frac{E_{\vis(3)}}{\vert\bm{p}^{\vis(3)}\vert \, z_{3}^{2}} \, \frac{1}{2 \, \hat{E}_{\vis(3)}} 
  \cdot \frac{\vert\mathcal{M}^{\eff(4)}_{\Pgt \to \tauh\Pnut}\vert^{2}}{256\pi^{6}} 
  \frac{E_{\vis(4)}}{\vert\bm{p}^{\vis(4)}\vert \, z_{4}^{2}} \, \frac{d\pThat^{\vis(4)}}{2 \, \hat{E}_{\vis(4)}} 
  \cdot W_{\Phadron}( \pT^{\vis(4)} | \pThat^{\vis(4)} ) \cdot \nonumber \\
& \qquad W_{\rec}( \pX^{\rec},\pY^{\rec} | \pXhat^{\rec},\pYhat^{\rec} ) 
\label{eq:likelihood_lllth}
\end{align}

\subsubsection{$\PHiggs\PHiggs \to \Pgt\Pgt\Pgt\Pgt \to \Plepton\Plepton\Plepton\Plepton$ decay channel}

\begin{align}
&
\mathcal{P}(\bm{p}^{\vis(1)},\bm{p}^{\vis(2)},\bm{p}^{\vis(3)},\bm{p}^{\vis(4)};\pX^{\rec},\pY^{\rec}|m_{\textrm{X}})
= \frac{32\pi^{8}}{m_{\Pgt} \, \Gamma_{\Pgt} \, s} \, \nonumber \\
& \qquad \int \, dz_{1} \, dm^{2}_{\inv(1)} \, d\phi_{\inv(1)} \, dm^{2}_{\inv(2)} \, d\phi_{\inv(2)} \, dm^{2}_{\inv(3)} \, d\phi_{\inv(3)} \, dm^{2}_{\inv(4)} \, d\phi_{\inv(4)} \, 
  \sum_{z_{3}^{+},z_{3}^{-}} \, \Bigr\lvert \frac{z_{1} \, z_{3}^{2}}{b \, z_{3}^{2} - c} \Bigr\rvert \cdot \nonumber \\
& \qquad \frac{I_{\inv(1)}}{512\pi^{6}} 
  \frac{E_{\vis(1)}}{\vert\bm{p}^{\vis(1)}\vert \, z_{1}^{2}} \, \frac{1}{2 \, \hat{E}_{\vis(1)}} 
  \cdot \frac{I_{\inv(2)}}{512\pi^{6}} 
  \frac{E_{\vis(2)}}{\vert\bm{p}^{\vis(2)}\vert \, z_{2}^{2}} \, \frac{1}{2 \, \hat{E}_{\vis(2)}} \cdot \nonumber \\
& \qquad \frac{I_{\inv(3)}}{512\pi^{6}} 
  \frac{E_{\vis(3)}}{\vert\bm{p}^{\vis(3)}\vert \, z_{3}^{2}} \, \frac{1}{2 \, \hat{E}_{\vis(3)}} 
  \cdot \frac{I_{\inv(4)}}{512\pi^{6}} 
  \frac{E_{\vis(4)}}{\vert\bm{p}^{\vis(4)}\vert \, z_{4}^{2}} \, \frac{1}{2 \, \hat{E}_{\vis(4)}} \cdot \nonumber \\
& \qquad W_{\rec}( \pX^{\rec},\pY^{\rec} | \pXhat^{\rec},\pYhat^{\rec} ) 
\label{eq:likelihood_llll}
\end{align}

%% file: svFit4tau.bbl
\begin{thebibliography}{10}
\expandafter\ifx\csname url\endcsname\relax
  \def\url#1{\texttt{#1}}\fi
\expandafter\ifx\csname urlprefix\endcsname\relax\def\urlprefix{URL }\fi
\expandafter\ifx\csname href\endcsname\relax
  \def\href#1#2{#2} \def\path#1{#1}\fi

\bibitem{Higgs-Discovery_CMS}
S.~Chatrchyan, et~al., {Observation of a new boson at a mass of 125 GeV with
  the CMS experiment at the LHC}, Phys.~Lett. B~716 (2012) 30.
\newblock \href {http://arxiv.org/abs/1207.7235} {\path{arXiv:1207.7235}},
  \href {http://dx.doi.org/10.1016/j.physletb.2012.08.021}
  {\path{doi:10.1016/j.physletb.2012.08.021}}.

\bibitem{Higgs-Discovery_ATLAS}
G.~Aad, et~al., {Observation of a new particle in the search for the Standard
  Model Higgs boson with the ATLAS detector at the LHC}, Phys.~Lett. B~716
  (2012) 1.
\newblock \href {http://arxiv.org/abs/1207.7214} {\path{arXiv:1207.7214}},
  \href {http://dx.doi.org/10.1016/j.physletb.2012.08.020}
  {\path{doi:10.1016/j.physletb.2012.08.020}}.

\bibitem{HIG-14-042}
G.~Aad, et~al., {Combined measurement of the Higgs boson mass in
  $\textrm{p}\textrm{p}$ collisions at $\sqrt{s}=7$ and $8$~TeV with the ATLAS
  and CMS experiments}, Phys.~Rev.~Lett. 114 (2015) 191803.
\newblock \href {http://arxiv.org/abs/1503.07589} {\path{arXiv:1503.07589}},
  \href {http://dx.doi.org/10.1103/PhysRevLett.114.191803}
  {\path{doi:10.1103/PhysRevLett.114.191803}}.

\bibitem{PDG}
C.~Patrignani, et~al., {Review of particle physics}, Chin.~Phys. C~40 (2016)
  100001.
\newblock \href {http://dx.doi.org/10.1088/1674-1137/40/10/100001}
  {\path{doi:10.1088/1674-1137/40/10/100001}}.

\bibitem{HIG-15-002}
G.~Aad, et~al., {Measurements of the Higgs boson production and decay rates and
  constraints on its couplings from a combined ATLAS and CMS analysis of the
  LHC $\textrm{p}\textrm{p}$ collision data at $\sqrt{s}=7$ and $8$~TeV}, JHEP
  08 (2016) 045.
\newblock \href {http://arxiv.org/abs/1606.02266} {\path{arXiv:1606.02266}},
  \href {http://dx.doi.org/10.1007/JHEP08(2016)045}
  {\path{doi:10.1007/JHEP08(2016)045}}.

\bibitem{HIG-13-004}
S.~Chatrchyan, et~al., {Evidence for the $125$~GeV Higgs boson decaying to a
  pair of $\tau$ leptons}, JHEP 05 (2014) 104.
\newblock \href {http://arxiv.org/abs/1401.5041} {\path{arXiv:1401.5041}},
  \href {http://dx.doi.org/10.1007/JHEP05(2014)104}
  {\path{doi:10.1007/JHEP05(2014)104}}.

\bibitem{Aad:2015vsa}
G.~Aad, et~al., {Evidence for the Higgs-boson Yukawa coupling to $\tau$ leptons
  with the ATLAS detector}, JHEP 04 (2015) 117.
\newblock \href {http://arxiv.org/abs/1501.04943} {\path{arXiv:1501.04943}},
  \href {http://dx.doi.org/10.1007/JHEP04(2015)117}
  {\path{doi:10.1007/JHEP04(2015)117}}.

\bibitem{HIG-16-043}
A.~M. Sirunyan, et~al., {Observation of the Higgs boson decay to a pair of
  $\tau$ leptons with the CMS detector}, Phys.~Lett. B~779 (2018) 283.
\newblock \href {http://arxiv.org/abs/1708.00373} {\path{arXiv:1708.00373}},
  \href {http://dx.doi.org/10.1016/j.physletb.2018.02.004}
  {\path{doi:10.1016/j.physletb.2018.02.004}}.

\bibitem{ATLAS:2018lur}
{ATLAS Collaboration}, \href{https://cds.cern.ch/record/2621794}{{Cross-section
  measurements of the Higgs boson decaying to a pair of $\tau$ leptons in
  $\textrm{p}\textrm{p}$ collisions at $\sqrt{s}=13$~TeV with the ATLAS
  detector}}, ATLAS Note ATLAS-CONF-2018-021.
\newline\urlprefix\url{https://cds.cern.ch/record/2621794}

\bibitem{Branco:2011iw}
G.~C. Branco, P.~M. Ferreira, L.~Lavoura, M.~N. Rebelo, M.~Sher, J.~P. Silva,
  {Theory and phenomenology of two-Higgs-doublet models}, Phys.~Rept. 516
  (2012) 1.
\newblock \href {http://arxiv.org/abs/1106.0034} {\path{arXiv:1106.0034}},
  \href {http://dx.doi.org/10.1016/j.physrep.2012.02.002}
  {\path{doi:10.1016/j.physrep.2012.02.002}}.

\bibitem{Gunion:1989we}
J.~F. Gunion, H.~E. Haber, G.~L. Kane, S.~Dawson, {The Higgs hunter's guide},
  Front.~Phys. 80 (2000) 1, [Erratum: arXiv:hep-ph/9302272~(1992)].

\bibitem{Grober:2010yv}
R.~Grober, M.~{M\"{u}hlleitner}, {Composite Higgs boson pair production at the
  LHC}, JHEP 06 (2011) 020.
\newblock \href {http://arxiv.org/abs/1012.1562} {\path{arXiv:1012.1562}},
  \href {http://dx.doi.org/10.1007/JHEP06(2011)020}
  {\path{doi:10.1007/JHEP06(2011)020}}.

\bibitem{Contino:2012xk}
R.~Contino, M.~Ghezzi, M.~Moretti, G.~Panico, F.~Piccinini, A.~Wulzer,
  {Anomalous couplings in double Higgs production}, JHEP 08 (2012) 154.
\newblock \href {http://arxiv.org/abs/1205.5444} {\path{arXiv:1205.5444}},
  \href {http://dx.doi.org/10.1007/JHEP08(2012)154}
  {\path{doi:10.1007/JHEP08(2012)154}}.

\bibitem{Baur:2002rb}
U.~Baur, T.~Plehn, D.~L. Rainwater, {Measuring the Higgs boson self coupling at
  the LHC and finite top mass matrix elements}, Phys.~Rev.~Lett. 89 (2002)
  151801.
\newblock \href {http://arxiv.org/abs/hep-ph/0206024}
  {\path{arXiv:hep-ph/0206024}}, \href
  {http://dx.doi.org/10.1103/PhysRevLett.89.151801}
  {\path{doi:10.1103/PhysRevLett.89.151801}}.

\bibitem{Baur:2002qd}
U.~Baur, T.~Plehn, D.~L. Rainwater, {Determining the Higgs boson self coupling
  at hadron colliders}, Phys.~Rev. D~67 (2003) 033003.
\newblock \href {http://arxiv.org/abs/hep-ph/0211224}
  {\path{arXiv:hep-ph/0211224}}, \href
  {http://dx.doi.org/10.1103/PhysRevD.67.033003}
  {\path{doi:10.1103/PhysRevD.67.033003}}.

\bibitem{Englert:2011yb}
C.~Englert, T.~Plehn, D.~Zerwas, P.~M. Zerwas, {Exploring the Higgs portal},
  Phys.~Lett. B~703 (2011) 298.
\newblock \href {http://arxiv.org/abs/1106.3097} {\path{arXiv:1106.3097}},
  \href {http://dx.doi.org/10.1016/j.physletb.2011.08.002}
  {\path{doi:10.1016/j.physletb.2011.08.002}}.

\bibitem{No:2013wsa}
J.~M. No, M.~Ramsey-Musolf, Probing the {H}iggs portal at the {LHC} through
  resonant di-{H}iggs production, Phys.~Rev. D~89 (2014) 095031.
\newblock \href {http://arxiv.org/abs/1310.6035} {\path{arXiv:1310.6035}},
  \href {http://dx.doi.org/10.1103/PhysRevD.89.095031}
  {\path{doi:10.1103/PhysRevD.89.095031}}.

\bibitem{Randall:1999ee}
L.~Randall, R.~Sundrum, {A large mass hierarchy from a small extra dimension},
  Phys.~Rev.~Lett. 83 (1999) 3370.
\newblock \href {http://arxiv.org/abs/hep-ph/9905221}
  {\path{arXiv:hep-ph/9905221}}, \href
  {http://dx.doi.org/10.1103/PhysRevLett.83.3370}
  {\path{doi:10.1103/PhysRevLett.83.3370}}.

\bibitem{SVfitMEM}
L.~Bianchini, B.~Calpas, J.~Conway, A.~Fowlie, L.~Marzola, C.~Veelken,
  L.~Perrini, {Reconstruction of the Higgs mass in events with Higgs bosons
  decaying into a pair of $\tau$ leptons using matrix element techniques},
  Nucl.~Instrum.~Meth. A~862 (2017) 54.
\newblock \href {http://arxiv.org/abs/1603.05910} {\path{arXiv:1603.05910}},
  \href {http://dx.doi.org/10.1016/j.nima.2017.05.001}
  {\path{doi:10.1016/j.nima.2017.05.001}}.

\bibitem{Kondo:1988yd}
K.~Kondo, {Dynamical likelihood method for reconstruction of events with
  missing momentum. 1: method and toy models}, J.~Phys.~Soc.~Jap. 57 (1988)
  4126--4140.
\newblock \href {http://dx.doi.org/10.1143/JPSJ.57.4126}
  {\path{doi:10.1143/JPSJ.57.4126}}.

\bibitem{Kondo:1991dw}
K.~Kondo, {Dynamical likelihood method for reconstruction of events with
  missing momentum. 2: mass spectra for $2 \to 2$ processes},
  J.~Phys.~Soc.~Jap. 60 (1991) 836--844.
\newblock \href {http://dx.doi.org/10.1143/JPSJ.60.836}
  {\path{doi:10.1143/JPSJ.60.836}}.

\bibitem{HIG-14-002}
V.~Khachatryan, et~al., {Constraints on the Higgs boson width from off-shell
  production and decay to $\textrm{Z}$-boson pairs}, Phys.~Lett. B~736 (2014)
  64.
\newblock \href {http://arxiv.org/abs/1405.3455} {\path{arXiv:1405.3455}},
  \href {http://dx.doi.org/10.1016/j.physletb.2014.06.077}
  {\path{doi:10.1016/j.physletb.2014.06.077}}.

\bibitem{Aad:2015xua}
G.~Aad, et~al., {Constraints on the off-shell Higgs boson signal strength in
  the high-mass $\textrm{Z}\textrm{Z}$ and $\textrm{W}\textrm{W}$ final states
  with the ATLAS detector}, Eur.~Phys.~J. C~75~(7) (2015) 335.
\newblock \href {http://arxiv.org/abs/1503.01060} {\path{arXiv:1503.01060}},
  \href {http://dx.doi.org/10.1140/epjc/s10052-015-3542-2}
  {\path{doi:10.1140/epjc/s10052-015-3542-2}}.

\bibitem{VAMP}
T.~Ohl, {Vegas revisited: Adaptive Monte Carlo integration beyond
  factorization}, Comput.~Phys.~Commun. 120 (1999) 13.
\newblock \href {http://arxiv.org/abs/hep-ph/9806432}
  {\path{arXiv:hep-ph/9806432}}, \href
  {http://dx.doi.org/10.1016/S0010-4655(99)00209-X}
  {\path{doi:10.1016/S0010-4655(99)00209-X}}.

\bibitem{VEGAS}
G.~P. Lepage, {A new algorithm for adaptive multidimensional integration},
  J.~Comput.~Phys. 27 (1978) 192.
\newblock \href {http://dx.doi.org/10.1016/0021-9991(78)90004-9}
  {\path{doi:10.1016/0021-9991(78)90004-9}}.

\bibitem{Metropolis_Hastings}
W.~K. Hastings, {Monte Carlo sampling methods using Markov Chains and their
  applications}, Biometrika 57 (1970) 97.
\newblock \href {http://dx.doi.org/10.1093/biomet/57.1.97}
  {\path{doi:10.1093/biomet/57.1.97}}.

\bibitem{MadGraph_aMCatNLO}
J.~Alwall, R.~Frederix, S.~Frixione, V.~Hirschi, F.~Maltoni, O.~Mattelaer,
  H.~S. Shao, T.~Stelzer, P.~Torrielli, M.~Zaro, {The automated computation of
  tree-level and next-to-leading order differential cross sections, and their
  matching to parton shower simulations}, JHEP 07 (2014) 079.
\newblock \href {http://arxiv.org/abs/1405.0301} {\path{arXiv:1405.0301}},
  \href {http://dx.doi.org/10.1007/JHEP07(2014)079}
  {\path{doi:10.1007/JHEP07(2014)079}}.

\bibitem{NNPDF1}
R.~D. Ball, V.~Bertone, S.~Carrazza, L.~Del~Debbio, S.~Forte, A.~Guffanti,
  N.~P. Hartland, J.~Rojo, {Parton distributions with QED corrections},
  Nucl.~Phys. B~877 (2013) 290.
\newblock \href {http://arxiv.org/abs/1308.0598} {\path{arXiv:1308.0598}},
  \href {http://dx.doi.org/10.1016/j.nuclphysb.2013.10.010}
  {\path{doi:10.1016/j.nuclphysb.2013.10.010}}.

\bibitem{NNPDF2}
R.~D. Ball, V.~Bertone, F.~Cerutti, L.~Del~Debbio, S.~Forte, A.~Guffanti, J.~I.
  Latorre, J.~Rojo, M.~Ubiali, {Unbiased global determination of parton
  distributions and their uncertainties at NNLO and at LO}, Nucl.~Phys. B~855
  (2012) 153.
\newblock \href {http://arxiv.org/abs/1107.2652} {\path{arXiv:1107.2652}},
  \href {http://dx.doi.org/10.1016/j.nuclphysb.2011.09.024}
  {\path{doi:10.1016/j.nuclphysb.2011.09.024}}.

\bibitem{NNPDF3}
R.~D. Ball, et~al., {Parton distributions for the LHC Run II}, JHEP 04 (2015)
  040.
\newblock \href {http://arxiv.org/abs/1410.8849} {\path{arXiv:1410.8849}},
  \href {http://dx.doi.org/10.1007/JHEP04(2015)040}
  {\path{doi:10.1007/JHEP04(2015)040}}.

\bibitem{pythia8}
T.~Sjostrand, S.~Mrenna, P.~Z. Skands, {A brief introduction to PYTHIA 8.1},
  Comput.~Phys.~Commun. 178 (2008) 852.
\newblock \href {http://arxiv.org/abs/0710.3820} {\path{arXiv:0710.3820}},
  \href {http://dx.doi.org/10.1016/j.cpc.2008.01.036}
  {\path{doi:10.1016/j.cpc.2008.01.036}}.

\bibitem{PYTHIA_CUETP8M1tune_CMS}
V.~Khachatryan, et~al., {Event generator tunes obtained from underlying event
  and multiparton scattering measurements. }, Eur.~Phys.~J. C~76 (2016) 155.
\newblock \href {http://arxiv.org/abs/1512.00815} {\path{arXiv:1512.00815}},
  \href {http://dx.doi.org/10.1140/epjc/s10052-016-3988-x}
  {\path{doi:10.1140/epjc/s10052-016-3988-x}}.

\end{thebibliography}
